\documentclass[12pt,a4paper]{article}
\usepackage[T1]{fontenc}    % use a scalable T1 font encoding
\usepackage{lmodern}        % Latin Modern is the scalable successor to CM
\usepackage[numbers,sort&compress]{natbib}
\usepackage{amsmath,amssymb,tabularx}
\usepackage[title]{appendix}
\usepackage{verbatim}
\usepackage{titlesec}
\titlelabel{\thetitle.\quad}
\usepackage{bm}
\usepackage{subfigure}
\usepackage{dsfont}
\usepackage{color}
\usepackage{graphicx}
\usepackage{physics}
\usepackage{slashed}
\usepackage[svgnames]{xcolor}
\usepackage{hyperref}
\usepackage{hyperref}
\hypersetup{
	colorlinks,
	citecolor=blue,
	filecolor=blue,
	linkcolor=blue,
	urlcolor=blue
}
\usepackage{caption}	
\usepackage{natbib}
\usepackage{microtype}
\usepackage{ctable}
\usepackage{braket}

\newcommand\varpm{\mathbin{\vcenter{\hbox{%
				\oalign{\hfil$\scriptstyle+$\hfil\cr
					\noalign{\kern-.3ex}
					$\scriptscriptstyle({-})$\cr}%
}}}}
\newcommand\varmp{\mathbin{\vcenter{\hbox{%
				\oalign{$\scriptstyle({+})$\cr
					\noalign{\kern-.3ex}
					\hfil$\scriptscriptstyle-$\hfil\cr}%
}}}}
\newcommand{\nn}{\nonumber}
\usepackage{epsfig}
\usepackage{here}
\usepackage{mathrsfs}
\allowdisplaybreaks
\newcommand{\ba}{\begin{align}}
\newcommand{\ea}{\end{align}}
\graphicspath{{./Figures/}}
\def\nn{\nonumber}
\def\beq{\begin{equation}}
	\def\eeq{\end{equation}}
\def\bea{\begin{eqnarray}}
	\def\eea{\end{eqnarray}}
\begin{document}
\title{
\begin{flushright}
\ \\*[-80pt] 
\begin{minipage}{0.2\linewidth}
\normalsize

%arXiv:YYMM.NNNN \\
HUPD-2406 \\*[5pt]
\end{minipage}
\end{flushright}
{\Large 
$B_{(s)}\to D_{(s)}^{(*)}M$ decays in the presence of\\
final-state interaction\\*[5pt]} }
\author{\normalsize
\centerline{
Albertus Hariwangsa Panuluh$^{1,2}$\footnote{panuluh@usd.ac.id},
Satoshi Tanaka$^{3}$\footnote{tanaka.satoshi.college@gmail.com}, Hiroyuki Umeeda$^{4,5,6}$\footnote{umeeda@jlu.edu.cn}
} \\ \normalsize
\centerline{
%Author 3 $^{1}$\footnote{E-mail address},
%Author 4$^{1}$\footnote{E-mail address},
}
%%%%%%%%%%%%%%%%%%%%%%
\\*[10pt]
\centerline{
\begin{minipage}{\linewidth}
\begin{center}
$^1${\it \small
		Department of Physics Education, Faculty of Teacher Training and Education,\\ Sanata Dharma University,
		Paingan, Maguwohardjo, Sleman, Yogyakarta 55282, Indonesia} \\*[5pt]
$^2${\it \small
Physics Program, Graduate School of Advanced Science and Engineering, Hiroshima University,\\1-3-1 Kagamiyama, Higashi-Hiroshima 739-8526, Japan} \\*[5pt]
$^3${\it \small
Ibaraki prefecture, Japan} \\*[5pt]
$^4${\it \small
	Center for Theoretical Physics and College of Physics, Jilin University,\\
	Changchun, 130012, China} \\*[5pt]
$^5${\it \small
		China Center of Advanced Science and Technology,\\
		Beijing 100190, China} \\*[5pt]
$^6${\it \small
	Institute of High Energy Physics, Chinese Academy of Sciences,\\
	Beijing 100049, China} \\*[5pt]
\end{center}
\end{minipage}}
\\*[50pt]}
\date{
\centerline{\small \bf Abstract}
\begin{minipage}{0.9\linewidth}
\medskip
\medskip
\small
In light of the recent data for $\bar{B}_{(s)}\to D^{(*)}_{(s)}P$ and $\bar{B}_{(s)}\to D_{(s)}V$ decays, we perform a model-independent phenomenological analysis in the presence of quasi-elastic rescattering. With the Wilson coefficients including contributions beyond the standard model, lifetimes of $B$ meson as well as the $B^0_d-\bar{B}^0_d$ mixing are investigated for clarifying correlations among the observables. We show that parameter regions for quasi-elastic rescattering, the size of color-suppressed tree amplitudes and new physics are constrained due to the lifetime data. As a consequence, it is revealed that this scenario can be testable by the future LHCb measurement of width difference in $B^0_d-\bar{B}^0_d$ mixing and semi-leptonic CP asymmetry. 
\end{minipage}
}
\begin{titlepage}
\maketitle
\thispagestyle{empty}
\end{titlepage}
\counterwithin*{equation}{section}
\renewcommand\theequation{\thesection.\arabic{equation}}
\section{Introduction}\label{Sec:1}
Decays of $B$ mesons played an important role in testing the standard model (SM), as well as possible new physics (NP) contributions. Of the specific decay modes, nonleptonic channels are rather challenging processes in the context of strong interactions. A theoretical framework for these decays can be given by the QCD factorization  (QCDF) approach \cite{Beneke:2000ry}. In particular, it has been shown that for decays into heavy-light final states such as $\bar{B}_d\to D^+\pi^-$, vertex corrections are dominated by hard gluon exchange for large $m_b$ (see Ref.~\cite{Bauer:2001cu} for the factorization proof in the soft-collinear effective theory). Furthermore, there exist no penguin or annihilation diagrams for the mentioned channel. Owing to this observation, $\bar{B}_d\to D^+\pi^-$ decay is theoretically more tractable than those for light-light final states.
\par
Recently, it was pointed out \cite{Bordone:2020gao} that there are discrepancies between the experimental data\footnote{See Ref.~\cite{Belle:2022afp} for the recent experimental result. As to the theoretical side, recent discussion for $B\to DP$ decays in regards to SU(3) breaking is found in Ref.~\cite{Davies:2024vmv}.} \cite{ParticleDataGroup:2024cfk} and the prediction of the QCDF approach, where the theoretical analysis is performed at next-to-next-to-leading order (NNLO) \cite{Huber:2016xod}. It was also found that subleading power corrections, such as that from the three-particle Fock state of the light meson, {\it etc.}, are not large enough to explain the data.\footnote{In another recent work \cite{Piscopo:2023opf}, the analysis was carried out in light-cone QCD sum rules, giving an alternative prediction to the QCDF. While explaining the data within uncertainty, it was commented \cite{Piscopo:2023opf} that the additional investigations are required in view of the limited precision in the nonperturbative input.}
The mentioned circumstance possibly implies that final-state interactions (FSIs) \cite{Blok:1996uf, Blok:1997yj} are required for the nonleptonic decays, and/or NP contributions are present.
\par
In previous works, FSIs were discussed in the Regge theory \cite{Donoghue:1996hz} and addressed in the QCDF approach \cite{Beneke:2000ry}. A phenomenological framework incorporating FSIs was given by the quasielastic rescattering discussed in Refs~\cite{Chua:2001br, Chua:2005dt, Chua:2007qw, Chua:2018ikx}: in the limit of SU(3) symmetry, where mesons in the same flavor multiplet degenerate, FSIs are given by a mixing matrix that acts on the amplitudes with specific final states. An observable effect is a change in the relative phase between the amplitudes with final states lying in different SU(3) multiplets, since mixing between states with different quantum numbers does not occur and thus it only alters the phases. Formulated in this way, the quasielastic rescattering gives a tractable approach for including two-body FSIs.
\par
In Ref.~\cite{Endo:2021ifc}, it was shown that, even if the quasielastic rescattering is incorporated, the puzzle for the branching ratios cannot be resolved in a reasonable way, in the sense that color-allowed and color-suppressed processes are not simultaneously explained, with an overall coefficient of the color-suppressed tree amplitude treated as a free parameter. In this circumstance, the possibility that NP is affecting the short-distance Wilson coefficients is not straightforwardly ruled out and was investigated \cite{Endo:2021ifc} with the FSIs, where parameter regions are more extended as compared to the case without rescattering. See Refs.~\cite{Cai:2021mlt, Iguro:2020ndk, Fleischer:2021cct, Fleischer:2021cwb} for further studies in the context beyond the SM.
\par
It is worth noting that the aforementioned scenario with NP is supposed to confront constraints from other observables with nonleptonic transitions. This was pointed out in Ref.~\cite{Bordone:2020gao} (see also Ref.~\cite{Gershon:2021pnc}), while dedicated numerical results were obtained in Ref. \cite{Lenz:2022pgw}. In particular, the total widths of the $B$ meson and $B^0_d-\bar{B}^0_d$ mixing (see Refs.~\cite{Lenz:2022rbq,Lenz:2019lvd} for recent analyses) are considered as constraints on the NP scenario. For the former, a lifetime ratio $\tau(B^+)/\tau(B_d)$ plays a particularly suitable role, since theoretical uncertainty is better controlled and is characterized by the contribution of Pauli interference.
\par
In this work, we carry out a phenomenological analysis of $B_{(s)}\to D_{(s)}^{(*)}M$ in the presence of the quasielastic rescattering and clarify its correlation with $\tau(B^+)/\tau(B_d)$ and $B^0_d-\bar{B}^0_d$ mixing. We show that these observables lead to constraints and/or predictions of the scenario in which rescattering contributions are involved in $B_{(s)}\to D_{(s)}^{(*)}M$ decays. In particular, it is demonstrated that some of the  model-parameter space is significantly constrained to explain the observables. As a resulting prediction, the width difference $(\Delta \Gamma_d)$ and the semileptonic \textit{CP} asymmetry $(\mathcal{A}_{\rm SL}^d)$ are evaluated.
\par
This paper is organized as follows. In Sec.~\ref{Sec:2}, a basic framework for  quasielastic rescattering is introduced for $B\to DM$ decays. The constraints from branching ratios on the model parameters are obtained in an analytical manner, for both $b\to c\bar{u}s$ and $b\to c\bar{u}d$ transitions. The SU(3) symmetry breaking is considered within the formalism for the latter processes. In Secs.~\ref{Sec:3} and \ref{Sec:4}, $B$-meson lifetimes and $B^0-\bar{B}^0$ mixing are respectively discussed. In Sec.~\ref{Sec:5}, the phenomenological analysis is given for the mentioned observables. The correlation patterns for QCD factorization parameters and the rescattering angle  satisfying the phenomenological constraints are obtained numerically. We show that this scenario can be testable via 
$\Delta \Gamma_d$ and $\mathcal{A}_{\rm SL}^d$ with future LHCb measurements. Finally, concluding remarks are given in Sec.~\ref{Sec:6}.
%====================
\section{$B\to DM$ decays} \label{Sec:2}
%====================
In this section, we investigate $B$-meson nonleptonic decays into two-body exclusive final states that include a charmed meson. The effective Hamiltonian relevant for $b\to c\bar{q}_2q_3~( q_2=u, c, q_3=d, s)$ is given by,
\bea
\mathcal{H}_W&=&\frac{G_F}{\sqrt{2}}\left[V_{c b}V_{q_2q_3}^*\displaystyle\sum_{i=1}^{2}c_iQ_i^{\bar{q}_2q_3}
-V_{t b}V_{tq_3}^*\left(\displaystyle\sum_{i=3}^{6}c_iQ_i^{q_3}+c_{8}Q_{8}^{q_3}\right)
\right].\label{Eq:BurasHw}
\eea
The definitions of the operators that appear in Eq.~(\ref{Eq:BurasHw}) are given in Eq.~(\ref{App:A1}). The radiative QCD corrections to the Wilson coefficient can be obtained in Ref.~\cite{Buchalla:1995vs} and references therein, with a certain care of the difference in the notation. 
\begin{comment}
	%====================
	\subsection{Topological amplitudes in QCDF \textcolor{red}{to be written}} \label{Sec:2:1}
	%====================
	In the QCDF approach \cite{Beneke:2000ry}, it has been shown that $B$-meson decays into heavy-light final states are calculated by the hard scattering in QCD 
	
	\bea
	a_1&=&\frac{N_c+1}{2N_c}\bar{C}_+(\mu)+\frac{N_c-1}{2N_c}\bar{C}_-(\mu)\nn\\
	&&+\frac{\alpha_s}{4\pi}\frac{C_F}{2N_c}C_8(\mu)\left[
	-6\ln\frac{\mu^2}{m_b^2}+\int_0^1du F(u, z)\Phi_L(u)\right]
	\eea
\end{comment}
%====================
\subsection{Quasielastic rescattering}\label{Sec:2:2}
%====================
Here we recapitulate the FSI discussed in Refs.~\cite{Chua:2001br, Chua:2005dt, Chua:2007qw, Chua:2018ikx}; see also Ref.~\cite{Endo:2021ifc}. Decay amplitudes without FSIs are given by vector notations and classified as $A_{S, I_z}$, where $S$ and $I_z$ denote
the strangeness and the diagonalized component of isospin,
\bea
\mathcal{A}_{-1, 0}&=&\begin{pmatrix}
	\mathcal{A}(\bar{B}^0\to D^+K^-)\\
	\mathcal{A}(\bar{B}^0\to D^0\bar{K}^0)
\end{pmatrix},\qquad
\mathcal{A}_{1, -1}=\begin{pmatrix}
	\mathcal{A}(\bar{B}^0_s\to D^+_s\pi^-)\\
	\mathcal{A}(\bar{B}^0_s\to D^0K^0)
\end{pmatrix}.\label{Eq:ampbefore}
\eea
The FSIs can be taken into account by the quasielastic scattering; due to $\bar{3}\times 8= \overline{15}+6+\bar{3}$ for the final state that consists of $D\Pi$, where $\Pi$ is an SU(3) octet state, the rescattering matrix is decomposed as \cite{Chua:2001br, Chua:2005dt, Chua:2007qw}
\bea
S^{1/2}=e^{i\delta_{\overline{15}}}\ket{\overline{15}; a}\bra{\overline{15}; a}
+e^{i\delta_{6}}\ket{6; b}\bra{6; b}
+\displaystyle\sum_{m, n=\overline{3}, \overline{3}^\prime}
\ket{m; c}\mathcal{U}_{mn}^{1/2}\bra{n; c}.\label{Eq:rescattering}
\eea
For the $\overline{15}$ and $6$ terms in Eq.~(\ref{Eq:rescattering}), in the limit of the flavor symmetry, the final states with definitive quantum numbers such as isospin do not mix under the FSIs and thus the rescattering merely alters the phase of the amplitude. In contrast to this case, for the last term in Eq.~(\ref{Eq:rescattering}), one needs to take account of the mixing between $\bar{3}$ and $\bar{3}^\prime$ states in the presence of the SU(3) singlet state that consists of light flavors accompanied by a $D$ meson. This is represented as $2\times 2$ matrix given by $\mathcal{U}_{mn}^{1/2}$ in Eq.~(\ref{Eq:rescattering}).\par
Incorporating the FSIs, the amplitudes in Eq.~(\ref{Eq:ampbefore}) are modified as
\bea
\mathcal{A}^{f}_{S, I_z}=V^{-1}_{S, I_z}S^{1/2}_{S, I_z}V_{S, I_z}\mathcal{A}_{S, I_z},
\eea
where $S^{1/2}_{S, I_z}$
represents the rescattering matrix for specific quantum numbers while $V_{S, I_z}$ is a diagonal matrix defined by \cite{Chua:2018ikx, Endo:2021ifc}
\bea
V_{-1, 0}=\mathrm{diag}\left(1, 1\right),\qquad
V_{1, -1}=\mathrm{diag}\left(1, \frac{f_{D_s}f_\pi}{f_Df_K}\right).\label{Eq:SU(3)breaking}
\eea
Due to Eq.~(\ref{Eq:SU(3)breaking}), SU(3) breaking for the rescattering is included in $b\to c\bar{u}d$ via the decay constants, but not in $b\to c\bar{u}s$. If we consider the state with $S=-1$ and $I_z=0$ as an example,
the rescattering matrix that mixes $D^+ K^-$ and $D^0\bar{K}^0$ final states can be obtained from components of the SU(3) representations,
\bea
\overline{15}~(S=-1, I=1)&:&~\frac{1}{\sqrt{2}}(\ket{D^+K^-}+\ket{D^0\bar{K}^0}),\\
6~(S=-1, I=0)&:&~\frac{1}{\sqrt{2}}(\ket{D^+K^-}-\ket{D^0\bar{K}^0}),
\eea
without antitriplet states. Likewise, the decomposition of $S=1, I_z=-1$ can also be obtained. The above relations are readily solved with respect to $\ket{D^+K^-}$ and $\ket{D^0\bar{K}^0}$. By acting the matrix in Eq.~(\ref{Eq:rescattering}) on those states for both $S=-1, I_z=0$ and $S=1, I_z=-1$, one can obtain \cite{Chua:2001br}
\bea
S^{1/2}_{-1, 0}=S^{1/2}_{1, -1}=\frac{e^{i\delta_{\overline{15}}}}{2}
\begin{pmatrix}
	1 + e^{i\delta^\prime} & 1 - e^{i\delta^\prime}\\
	1 - e^{i\delta^\prime} & 1 + e^{i\delta^\prime}
\end{pmatrix},\qquad 
\delta^\prime = \delta_{6}-\delta_{\overline{15}},
\eea
where the overall phase denoted by $\delta_{\overline{15}}$ cancels out when the branching ratios are calculated. It should be noted that for the above two choices of strangeness and isospin, the antitriplet term in Eq.~(\ref{Eq:rescattering}) is not involved in the discussion.
\par
In the following sections, we also discuss processes with final states of $S=0, I_z=3/2$ and $S=1, I_z=1$, corresponding, {\it e.g.}, to $B^+\to \bar{D}^0\pi^+$ and $B^+\to \bar{D}^0K^+$. These cases do not undergo the rescattering since there are no other decay channels that mix together. Hence, the rescattering is considered for the $S=-1, I_z=0$ and $S=1, I_z=-1$ cases (or their \textit{CP}-conjugate processes), individually.
%====================
\subsection{Branching ratios}
%====================
In this section, relations constraining parameters of the QCDF approach and rescattering from branching ratios of $B$-meson two-body decays are obtained. For definitiveness, the discussion of $\bar{B}\to D\bar{K}$, which proceeds via $b\to c\bar{u}s$, is given first. Subsequently, other processes with $b\to c\bar{u}d$ transitions are also analyzed. The resulting relations in Eqs.~(\ref{Eq:Sol1})-(\ref{Eq:Sol3}) and (\ref{Eq:ReaSU3})-(\ref{Eq:SU3breakingsol}) play a major role in the numerical analysis.\par
\subsubsection{$b\to c\bar{u}s$}
Below, $\bar{B}\to D\bar{K}$ with the final state that consists of two pseudoscalars is discussed first. In the presence of the rescattering, branching ratios of the nonleptonic decays are
\bea
\textrm{Br}^{ij}&\equiv&\mathrm{Br}[P\to M_1^iM_2^j] = \frac{\tau^{ij}p_{\rm cm}[P\to M_1^iM_2^j]}{8\pi m_P^2}|V_{cb}V_{us}^*|^2|\mathcal{A}_f[P\to M_1^iM_2^j]|^2,\label{Eq:branch1}
\eea
with $(i, j)=(+, -), (0, 0), (0, -)$ and $\tau^{ij}$ denoting a lifetime of the initial particle, which is $\tau(B^+), \tau(B_d)$ or $\tau(B_s)$.
In Eq.~(\ref{Eq:branch1}), $p_{\rm cm}$ is a momentum of either particle in the final state defined in the rest frame of the initial particle,
\bea
p_{\rm cm}[P\to M_1M_2]&=&\frac{1}{2m_P}\sqrt{[m_P^2-(m_{M_1}+m_{M_2})^2][m_P^2-(m_{M_1}-m_{M_2})^2]}.
\eea
In Eq.~(\ref{Eq:branch1}), the subscript $f$ represents the presence of FSIs.
\par
In the case without rescattering, the processes are represented by topological amplitudes,
\bea
\mathcal{A}^{+-}&\equiv&\mathcal{A}[\bar{B}^0\to D^{+}K^{-}]=T_{DK},\label{Eq:amp1}\\
\mathcal{A}^{00}&\equiv&\mathcal{A}[\bar{B}^0\to D^0\bar{K}^0]=C_{DK},\label{Eq:amp2}\\
\mathcal{A}^{0-}&\equiv&\mathcal{A}[B^- \to D^0K^-]=T_{DK}+C_{DK},\label{Eq:amp3}
\eea
where $T_{DK}$ and $C_{DK}$ are color-allowed and -suppressed tree diagrams, respectively. In the QCDF approach \cite{Beneke:2000ry}, these amplitudes are evaluated as
\bea
T_{DK}=N_{DK}^Ta_1,\qquad C_{DK}=N_{DK}^Ca_2^{\rm eff}.\label{Eq:TDK}
\eea
In the above relation, $N_{DK}^{T(C)}$ is a normalization factor that is a product of the Fermi constant, the decay constant, and the form factor defined in Eq.~(\ref{Eq:Nfactors}). For later convenience, we introduce the notation
\bea
\bar{a}_2=(N^C_{DK}a_2^{\rm eff})/(N^T_{DK}a_1).\label{Eq:a2def}
\eea
\par
By using the three relations for $(i, j)=(+, -), (0, 0),$ and $(0, -)$ in Eq.~(\ref{Eq:branch1}), one can determine $\textrm{Re}(\bar{a}_2), \textrm{Im}(\bar{a}_2)$ and $\delta^\prime$ with the branching ratio data and a given value of $a_1$. With the deriviation discussed in Appendix.~\ref{App:01},
the results read
\bea
\mathrm{Re}(\bar{a}_2)&=&
\frac{(\tau^{+-}/\tau^{0-})\mathrm{Br}^{0-}-\mathrm{Br}^{+-}
	-\mathrm{Br}^{00}}{2\mathcal{N}_{DK}},\label{Eq:Sol1}\\
\mathrm{Im}(\bar{a}_2)
&=&\pm\sqrt{\frac{\mathrm{Br}^{+-}
		+\mathrm{Br}^{00}}{\mathcal{N}_{DK}}
	-1-[\mathrm{Re}(\bar{a}_2)]^2},\label{Eq:Sol2}\\
\delta^\prime&=&
\mathrm{Arcsin}\left[\frac{\mathrm{Br}^{+-}
	-\mathrm{Br}^{00}}{\mathcal{N}_{DK}\sqrt{A^2_{DK}+B^2_{DK}}}\right] - \omega_{DK},\quad \nn\\
&&\pi-\mathrm{Arcsin}\left[\frac{\mathrm{Br}^{+-}-\mathrm{Br}^{00}}{\mathcal{N}_{DK}\sqrt{\bar{A}^2_{DK}+\bar{B}^2_{DK}}}\right] - \omega_{DK},\quad (\mathrm{mod}~2\pi)\quad\quad\label{Eq:Sol3}
\eea
where the definitions of $\mathcal{N}_{DK}, A_{DK}, B_{DK}$, and $\omega_{DK}$ are given in Appendix~\ref{App:01}. It should be noted that there are twofold ambiguities for $\delta^\prime$ and the sign of $\textrm{Im}~(\bar{a}_2)$. The solutions in Eqs.~(\ref{Eq:Sol1})-(\ref{Eq:Sol3}) exist only if the following conditions are satisfied:
\bea
&\mathcal{N}_{DK}\neq 0,&\label{Eq:condi1}\\
&\displaystyle\frac{\mathrm{Br}^{+-}+\mathrm{Br}^{00}}{\mathcal{N}_{DK}}-[\mathrm{Re}(\bar{a}_2)]^2\geq 1,&\label{Eq:condi2}\\
&-1\leq \displaystyle\frac{\mathrm{Br}^{+-}
	-\mathrm{Br}^{00}}{\mathcal{N}_{DK}\sqrt{A^2_{DK}+B^2_{DK}}}\leq1.&\label{Eq:condi3}
\eea
The above conditions follow from the deriviation procedure in Appendix~\ref{App:01}.
\par
In what follows, the cases of $\bar{B}\to D\bar{K}^*$ and $\bar{B}\to D^*\bar{K}$ decays are discussed to obtain relations similar to Eqs.~(\ref{Eq:Sol1})-(\ref{Eq:condi3}). For processes including a vector meson in the final state, a formula for branching ratios analogous to Eq.~(\ref{Eq:branch1}) is
\bea
\mathrm{Br}[P\to M_1^*M_2] &=& \frac{\tau_{P}p_{\rm cm}[P\to M_1^*M_2]}{8\pi m_{P}^2}|V_{cb}V_{us}^*|^2
\displaystyle\sum_{\epsilon}|\mathcal{A}_f[P\to M_1^*M_2]|^2,\label{Eq:branch2}\\
\mathrm{Br}[P\to M_1M_2^*] &=& \frac{\tau_{P}p_{\rm cm}[P\to M_1M_2^*]}{8\pi m_{P}^2}|V_{cb}V_{us}^*|^2
\displaystyle\sum_{\epsilon}|\mathcal{A}_f[P\to M_1M_2^*]|^2\label{Eq:branch3}.
\eea
For the amplitudes in Eqs.~(\ref{Eq:branch2}) and (\ref{Eq:branch3}), the  polarization is factored out as follows:
\bea
\mathcal{A}_f[P\to M_1^*M_2]&=& (\epsilon^*\cdot p_B)\bar{\mathcal{A}}_f[P\to M_1^*M_2],\nn\\
\mathcal{A}_f[P\to M_1M_2^*]&=& (\epsilon^*\cdot p_B)\bar{\mathcal{A}}_f[P\to M_1M_2^*].
\eea
By evaluating the polarization sum,
\bea
\displaystyle\sum_{\rm \epsilon}|\epsilon^*\cdot p_B|^2=\left(\frac{m_B}{m_V}p_{\rm cm}\right)^2,
\eea
the branching ratios in Eqs.~(\ref{Eq:branch2}) and (\ref{Eq:branch3}) are recast into the forms
\bea
\mathrm{Br}[P\to M_1^*M_2] &=& \frac{\tau_{P}p_{\rm cm}^3[P\to M_1^*M_2]}{8\pi m_{M_1^*}^2}|V_{cb}V_{us}^*|^2|\bar{\mathcal{A}}_f[P\to M_1^*M_2]|^2,\\
\mathrm{Br}[P\to M_1M_2^*] &=& \frac{\tau_{P}p_{\rm cm}^3[P\to M_1M_2^*]}{8\pi m_{M_2^*}^2}|V_{cb}V_{us}^*|^2|\bar{\mathcal{A}}_f[P\to M_1M_2^*]|^2.
\eea
\par
One can also obtain the resulting relations in Eqs.~(\ref{Eq:Sol1})-(\ref{Eq:condi3}) for $\bar{B}\to D\bar{K}^*$ and $\bar{B}\to D^*\bar{K}$ by simply replacing $D\to D^*$ and $K\to K^*$, respectively, with the proper replacement of data for the branching ratio on the rhs. The definitions of normalization factors for the case including a vector meson are given in Eqs.~(\ref{Eq:Nfactors2}) and (\ref{Eq:Nfactors3}).
\subsubsection{$b\to c\bar{u}d$}
By making some replacements in the previous discussions for $b\to c\bar{u}s$ decays, we can also obtain similar results for $b\to c\bar{u}d$ decays. In this case, nonvanishing SU(3) breaking for the rescattering in Eq.~(\ref{Eq:SU(3)breaking}) must be taken into account. In addition, mass differences in hadrons for normalization factors and phase space need to be consistently included unlike the case of $b\to c\bar{u}s$ decays, where the isospin symmetry relates the masses of the relevant particles.
The parameters relevant for SU(3) breaking are defined in Appendix~\ref{App:02}.
\par
For $b\to c\bar{u}d$, we introduce a normalized coefficient for color-suppressed tree diagram,
\bea
\bar{a}_2=(N_C^{D^0\bar{K}^0}a_2^{\rm eff})/(N_T^{D^+_s\pi^-}a_1).
\eea
The above object is not to be confused with the one for $b\to c\bar{u}s$ in Eq.~(\ref{Eq:a2def}).\par
In a way analogous to $b\to c\bar{u}s$ decays, solutions of the parameters for $b\to c\bar{u}d$ decays are
\bea
\mathrm{Re}(\bar{a}_2)&=&
\frac{(1+\Delta_{DP}^{(1)})\frac{\tau^{+-}}{\tau^{0-}}\mathrm{Br}^{0-}-(1+\Delta_{DP}^{(2)})
	\left[\mathrm{Br}^{+-}+(1+\Delta^{(3)}_{DP})
	\mathrm{Br}^{00}\right]}
{2\mathcal{N}_{D_s^+\pi^-}}-\frac{1}{2}(1+\Delta_{DP}^{(2)})
\Delta_{DP}^{(4)},\label{Eq:ReaSU3}\qquad\qquad\\
\mathrm{Im}(\bar{a}_2)
&=&\pm\sqrt{(1+\Delta_{DP}^{(5)})
	\left[\frac{\mathrm{Br}^{+-}+(1+\Delta^{(3)}_{DP})\mathrm{Br}^{00}}{\mathcal{N}_{D_s^+\pi^-}}-1\right]-[\mathrm{Re}(\bar{a}_2)]^2},\label{Eq:ImaSU3}\\
\delta^\prime&=&\mathrm{Arcsin}\left(\frac{\mathrm{Br}^{+-}-(1+\Delta^{(3)}_{DP})
	\mathrm{Br}^{00}}{\mathcal{N}_{D_s^+\pi^-}\sqrt{A_{DP}^2+B_{DP}^2}}\right)-\omega_{DP},\nn\\
&&\pi-\mathrm{Arcsin}\left(\frac{\mathrm{Br}^{+-}-(1+\Delta^{(3)}_{DP})\mathrm{Br}^{00}}{\mathcal{N}_{D_s^+\pi^-}\sqrt{A_{DP}^2+B_{DP}^2}}\right)-\omega_{DP}\quad(\mathrm{mod}~2\pi),\qquad\qquad\label{Eq:SU3breakingsol}
\eea 
where the definitions of $\Delta_{DP}^{(i)}~(i=1,\cdots, 5), \mathcal{N}_{D_s^+\pi^-}, \omega_{DP}, A_{DP}$ and $B_{DP}$ are given in Appendix~\ref{App:02}. It is found that the twofold ambiguities exist for Eqs.~(\ref{Eq:ImaSU3}) and (\ref{Eq:SU3breakingsol}) and $b\to c\bar{u}s$ decays. The solutions in Eqs.~(\ref{Eq:ReaSU3})-(\ref{Eq:SU3breakingsol}) exist only if the conditions given below are satisfied:
\bea
&{\cal N}_{D_s^+\pi^-}\neq 0,&\label{Eq:a1SU3}\\
&(1+\Delta_{DP}^{(5)})\left[\displaystyle\frac{\mathrm{Br}^{+-}+(1+\Delta_{DP}^{(3)})\mathrm{Br}^{00}}{\mathcal{N}_{D_s^+\pi^-}}-1\right]-[\mathrm{Re}(\bar{a}_2)]^2\geq 0,&\label{Eq:secSU3}\\
&-1\leq
\displaystyle\frac{\mathrm{Br}^{+-}-(1+\Delta_{DP}^{(3)})\mathrm{Br}^{00}}{\mathcal{N}_{D_s^+\pi^-}\sqrt{A_{DP}^2+B_{DP}^2}}\leq 1.\label{Eq:thiSU3}&
\eea
As shown in Eqs.~(\ref{Eq:SU3param1}) and (\ref{Eq:SU3param2}), $\Delta_{DP}^{(i)}$ vanishes in the SU(3) limit. Hence, the structures of Eqs.~(\ref{Eq:ReaSU3}-\ref{Eq:thiSU3}) for $b\to c\bar{u}d$ are reduced to those for $b\to c\bar{u}s$ in Eqs.~(\ref{Eq:Sol1})-(\ref{Eq:condi3}) in the SU(3) limit. It should be noted that the dependence on heavy-to-light form factors appears solely from $\Delta^{(2)}_{DP}$ in Eq.~(\ref{Eq:SU3param1}).
\par
For other $b\to c\bar{u}d$ decays including a vector meson, the result corresponding to $\bar{B}\to D^*P$ can be obtained by the replacement of $D\to D^*,$ while the one for $\bar{B}\to DV$ decay can be given by $P\to V, K\to K^*,$ and $\pi\to \rho$ in Eqs.~(\ref{Eq:ReaSU3})-(\ref{Eq:thiSU3}).
%====================
\section{Lifetimes of $B$ mesons} \label{Sec:3}
%====================
In this section, we recapitulate how the total width of beauty mesons is evaluated at leading order (LO) in QCD. This observable is analyzed by means of the heavy quark expansion (HQE): after the correlation functions are computed in the Euclidean domain, the expression is analytically continued to the Minkowski region, leading to the $1/m_b$ expansion for the observable. See Refs.~\cite{Lenz:2022rbq, Egner:2024lay}
for the recent works within the SM.
\par
We restrict ourselves to the isospin limit, where $\mu_\pi, \mu_G$ for $B_d$ are identical to those for $B^+$. With $q=u, d$ and $B_u=B^+$, the total width is written as
\bea
\Gamma(B_q)=\Gamma^{2\mathchar`-\mathrm{quark}}+\Gamma^{4\mathchar`-\mathrm{quark}}_q.\label{Eq:widthtotal}
\eea
The lifetime ratio is calculated from the above objects,
\bea
\frac{\tau(B^+)}{\tau(B_d)}=1-\frac{\Gamma(B^+)-\Gamma(B_d)}{\Gamma(B^+)}
= 1-\frac{\Gamma^{4\mathchar`-\mathrm{quark}}_u-\Gamma^{4\mathchar`-\mathrm{quark}}_d}{\Gamma(B^+)}.\label{Eq:lifetimeratio}
\eea
In the isospin limit for the matrix elements, $\tau(B^+)/\tau(B_d)-1$ is proportional to the spectator effect. In what follows, the two terms in Eq.~(\ref{Eq:widthtotal}) are discussed.
%====================
\subsection{Two-quark operators}
%====================
In the limit of the isospin symmetry, $\Gamma^{2\mathchar`-\mathrm{quark}}$ in Eq.~(\ref{Eq:widthtotal}) does not depend on the label of $q$. The contributions from two-quark operators in the above equation are classified by the nonleptonic and semileptonic pieces,
\bea
\Gamma^{2\mathchar`-\mathrm{quark}}=\displaystyle\sum_{q_2, q_3}\Gamma_{\rm NL}(b\to c\bar{q}_2 q_3)+\displaystyle\sum_{\ell}\Gamma_{\rm SL}(b\to c\ell \bar{\nu}),\label{Eq:2-quark}
\eea
where the summations are taken for all of the possible combinations with $q_2=u, c, q_3=d, s$ and $\ell= e, \mu, \tau$. It should be noted that the $b\to u$ transition, neglected in Eq.~(\ref{Eq:2-quark}), is Cabibbo suppressed, while larger contributions arise from $b\to c$. The partial widths that appear in Eq.~(\ref{Eq:2-quark}) are expanded by $1/m_b$, leading to
\bea
\Gamma_{\rm NL}(b\to c\bar{q}_2q_3)&=&\Gamma_0|V_{cb}V_{q_2q_3}^*|^2
\left(C_{\rm LP}^{c\bar{q}_2q_3}+C_\pi^{c\bar{q}_2q_3}\frac{\mu_\pi^2}{m_b^2}+C_G^{c\bar{q}_2q_3}\frac{\mu_G^2}{m_b^2}\right),\label{Eq:non}\\
\Gamma_{\rm SL}(b\to c\ell \bar{\nu})&=&\Gamma_0|V_{cb}|^2\left(C_{\rm LP}^{c\ell \bar{\nu}}
+C_\pi^{c\ell \bar{\nu}}\frac{\mu_\pi^2}{m_b^2}+C_G^{c\ell \bar{\nu}}\frac{\mu_G^2}{m_b^2}\right),\label{Eq:semi}
\eea
where $\Gamma_0=G_F^2m_b^5/(192\pi^3)$. The matrix elements of the two-quark operators, $\mu_\pi^2$ and $\mu_G^2$, are defined in Eq.~(\ref{Eq:HQETMATELE}). Furthermore, the nonleptonic coefficients in Eq.~(\ref{Eq:non}) stem from quadratic combinations of the $|\Delta B|=1$ Wilson coefficients,
\bea
C^{c\bar{q}_2q_3}_{I}&=&
3c_1^2\mathcal{C}_{I,\: 11}^{c\bar{q}_2q_3}
+2c_1c_2
\mathcal{C}_{I,\: 12}^{c\bar{q}_2q_3}
+3c_2^2\mathcal{C}_{I,\: 22}^{c\bar{q}_2q_3},\label{Eq:Cscal}
\eea
where $I=\mathrm{LP}, \pi, G$. In Eq.~(\ref{Eq:Cscal}), the contribution of NP is contained only in $c_1$ and $c_2$ while $\mathcal{C}_{I,\: ij}^{c\bar{q}_2q_3} ~(i, j=1, 2, I=\mathrm{LP}, \pi, G)$  can be obtained in previous works, {\it e.g.,} Ref.~\cite{Lenz:2014jha} and references therein. 
%====================
\subsection{Four-quark operators}\label{Sec:3:2}
%====================
The contribution of the spectator effect in Eq.~(\ref{Eq:widthtotal}) is rewritten as,
\bea
\Gamma^{4\mathchar`-\mathrm{quark}}_u =
\Gamma_{\rm int},\quad
\Gamma^{4\mathchar`-\mathrm{quark}}_d =
\Gamma_{\rm ann},
\label{Eq:6and7}
\eea
where int and ann represent the Pauli interference and weak annihilation, respectively. The above objects are proportional to the matrix elements of four-quark operators defined in Eqs.~(\ref{Eq:opeQ1})-(\ref{Eq:opeQ4}) and (\ref{Eq:Matele4quark1})-(\ref{Eq:Matele4quark4}). In the case of dimension-six contributions, the matrix elements can be obtained from Ref.~\cite{Kirk:2017juj}(see Ref.~\cite{Black:2025} for the updated values), while dimension-seven operators are evaluated via the vacuum insertion approximation, leading to \cite{Cheng:2018rkz}
\bea
\Gamma_{\mathrm{int}}&=&\frac{G_F^2m_b^2}{12\pi}|V_{cb}V_{ud}^*|^2
f_{B}^2m_{B}(1-z)^2\left\{(c_1^2 +c_2^2+6c_1c_2)\right.\nn\\
&&\left.\times
\left[B_1-\left(\frac{1+z}{1-z}+\frac{1}{2}\right)\left(\frac{m_{B}^2}{m_b^2}-1\right)\right]
+6(c_1^2+c_2^2)\epsilon_1\right\},\\
\Gamma_{\mathrm{ann}}&=&-\frac{G_F^2m_b^2}{12\pi}|V_{cb}V_{ud}^*|^2f_{B}^2m_{B}(1-z)^2\nn\\
&&\times\left\{
\left(\frac{c_1^2}{3}+2c_1c_2 +3c_2^2\right)
\left[
\left(1+\frac{z}{2}\right)B_1-\left(1+2z\right)B_2\right.
\right.\nn\\
&&+\left.\left[\frac{1+z+z^2}{1-z}+\frac{6z^2}{1-z}-\frac{1}{2}\left(1+\frac{z}{2}\right)-\frac{1}{2}(1+2z)\right]\left(\frac{m_{B}^2}{m_b^2}-1\right)\right]
\nn\\
&&\left.+2c_1^2
\left[\left(1+\frac{z}{2}\right)\epsilon_1-(1+2z)\epsilon_2\right]
\right\}.
\eea
In the above relations, $z$ represents $(m_c/m_b)^2$.
%====================
\section{$B^0_d-\bar{B}^0_d$ mixing} \label{Sec:4}
%====================
In this section, observables for neutral meson mixing of beauty mesons are discussed. In previous works, NP contributions to the width differences in the $D^0-\bar{D}^0$ and $B_s^0-\bar{B}_s^0$ mixings were discussed in Ref.~\cite{Golowich:2006gq} and Refs.~\cite{Badin:2007bv, Bobeth:2011st}. Moreover, \textit{CP} violation in the $B^0-\bar{B}^0$ mixing was also investigated beyond the SM in Refs.~\cite{Badin:2007bv,Trott:2010iz, Kim:2010gx, Nandi:2011uw, Dobrescu:2010rh, Oh:2010vc, Kubo:2010mh, Bai:2010kf,Altmannshofer:2011rm}.
\subsection{Dispersive part and absorptive part}\label{Sec:5:1}
%====================
The dispersive part for the $B^0_{d}-\bar{B}^0_{d}$ mixing amplitude in the SM is dominated by the contribution of intermediate top quarks. In this case, an expression where external quark momenta and masses are neglected, represented by the Inami-Lim function \cite{Inami:1980fz},
\bea
M_{21}&=&\frac{G_F^2M_W^2 }{12\pi^2}m_{B_d}f_{B_d}^2[\eta(\mu_b)]_{\rm VLL}B_1^d(\mu_b)S_0\left(\frac{\bar{m}_t^2(m_t)}{M_W^2}\right)(V_{tb}^*V_{td})^2,\label{Eq:M21}\\
S_0(x)&=&\frac{4x-11x^2+x^3}{4(1-x)^2}-\frac{3x^3\mathrm{ln}x}{2(1-x)^3},
\eea
gives an excellent approximation.\par
For the absorptive part, the theoretical analysis can be performed by HQE, 
analogously to the total width of $B$ mesons. In contrast to the case of the total width, the leading contribution to the width difference arises from four-quark operators. At next-to-leading order (NLO) in power corrections $(1/m_b)$, the width difference in the $B^0-\bar{B}^0$ mixing is obtained \cite{Beneke:1996gn,Beneke:1998sy, Ciuchini:2003ww}.\footnote{See also NNLO in the power correction $(1/m_b^2)$ in Ref.~\cite{Badin:2007bv}.} The SM contribution to $\Gamma_{21}$ in the $B_d^0-\bar{B}_d^0$ mixing with NLO power correction is given by \cite{Ciuchini:2003ww}
\bea
\Gamma_{21}=-\frac{G_F^2 m_b^2}{24\pi m_{B_d}}[c_1^{d, \mathrm{mix}}(\mu_2)\bra{\bar{B}_d^0}\mathcal{O}_1^d\ket{B_d^0}
+c_2^{d, \mathrm{mix}}(\mu_2)\bra{\bar{B}_d^0}\mathcal{O}_2^d\ket{B_d^0}+\delta_{1/m}^d].\label{Eq:Gamma21}
\eea
The expressions for the coefficients are given by \cite{Ciuchini:2003ww}
\bea
c_k^{d, \mathrm{mix}}&=&(V_{tb}^*V_{td})^2D_k^{uu}+
2V_{cb}^*V_{cd}V_{tb}^*V_{td}(D_k^{uu}-D_k^{cu})
+(V_{cb}^*V_{cd})^2(D_k^{uu}+D_k^{cc}-2D_k^{cu}),\quad (k=1, 2)\qquad\qquad\\
\delta^d_{1/m}&=&(V_{tb}^*V_{td})^2\delta^{uu\: d}_{1/m}+
2V_{cb}^*V_{cd}V_{tb}^*V_{td}(\delta^{uu\: d}_{1/m}-\delta^{cu\: d}_{1/m})+(V_{cb}^*V_{cd})^2(\delta^{uu\: d}_{1/m}+\delta^{cc\: d}_{1/m}-2\delta^{cu\: d}_{1/m}).
\eea
For $(q_1, q_2)=(c, c), (c, u)$ and $(u, u)$ with $k=1, 2$,
\bea
D_k^{q_1q_2}(\mu_2)&=&\displaystyle\sum_{i, j=1, 2}c_i(\mu_1)c_j(\mu_1)F^{q_1q_2,\mathrm{mix}}_{k, ij}(\mu_1, \mu_2)
+\frac{\alpha_s}{4\pi}[c_1(\mu_1)]^2P_{k, 11}^{q_1q_2}(\mu_1, \mu_2)\nn\\
&+&\frac{\alpha_s}{4\pi}c_1c_8(P_{k, 18}^{q_1}+P_{k, 18}^{q_2})+\displaystyle\sum_{i=1, 2}\displaystyle\sum_{r=3, 6}
c_ic_r(P_{k, ir}^{q_1}+P_{k, ir}^{q_2}).\label{Eq:Dresmix}
\eea
The phase-space functions were calculated in Refs.~\cite{Ciuchini:2003ww, Beneke:2003az} at the precision of NLO in QCD. It should be noted that in our notation of $c_1$ and $c_2$, we need to replace the indices $1\to 2$ and $2\to 1$ for $i, j$ in Refs.~\cite{Ciuchini:2003ww, Beneke:2003az}. The phase-space integral proportional to the quadratic term with respect to $c_1$ and $c_2$ is decomposed by the LO and NLO parts in QCD,
\bea
F^{q_1q_2,\mathrm{mix}}_{k, ij}=A^{q_1q_2,\mathrm{mix}}_{k, ij}+\frac{\alpha_s}{4\pi}B^{q_1q_2,\mathrm{mix}}_{k, ij}.\label{Eq:Fmix}
\eea
$A^{q_1q_2,\mathrm{mix}}_{k, ij}$ and $B^{q_1q_2,\mathrm{mix}}_{k, ij}$ in Eq.~(\ref{Eq:Fmix}), as well as the phase-space functions related to the penguin operators in Eq.~(\ref{Eq:Dresmix}), can be extracted from Ref.~\cite{Ciuchini:2003ww}
while $D^{cu,\mathrm{mix}}_k$ can be extracted from Ref.~\cite{Beneke:2003az}.
\par
The dimension-seven contributions were also obtained in Ref.~\cite{Ciuchini:2003ww},
\bea
\delta_{1/m}^{cc\:  d}&=&\sqrt{1-4z}\left\{(1+2z)[K_2(\braket{R_2^d}+2\braket{R_4^d})-2K_1(\braket{R_1^d}+\braket{R_2^d})]\right.\nn\\
&-&\left.\frac{12z^2}{1-4z}[K_1(\braket{R_2^d}+2\braket{R_3^d})+2K_2\braket{R_3^d}]\right\},\\
\delta_{1/m}^{cu\:  d}&=&(1-z)^2\left\{(1+2z)[K_2(\braket{R_2^d}+2\braket{R_4^d})-2K_1(\braket{R_1^d}+\braket{R_2^d})]\right.\nn\\
&-&\left.\frac{6z^2}{1-z}[K_1(\braket{R_2^d}+2\braket{R_3^d})+2K_2\braket{R_3^d}]\right\},\\
\delta_{1/m}^{uu\:  d}&=&K_2(\braket{R_2^d}+2\braket{R_4^d})-2K_1(\braket{R_1^d}+\braket{R_2^d}),
\eea
with $K_1=3c_2^2+2c_1c_2$ and $K_2=c_1^2$. The width difference in the $B_d$ system is given by \cite{Ciuchini:2003ww},
\bea
\Delta\Gamma_d=-2|M_{21}|\mathrm{Re}\left(\frac{\Gamma_{21}}{M_{21}}\right).\label{Eq:Ciucalc}
\eea
%=============
\subsection{\textit{CP} violation}
%=============
\textit{CP} violation in the $B^0_d-\bar{B}^0_d$ mixing can be measured in, {\it e.g.,} the semileptonic \textit{CP} asymmetry, given by
\bea
\mathcal{A}_{\rm SL}^d(t)=\frac{N[\bar{B}^0_d(t)\to \ell^+\nu_\ell X]-N[B_d(t)\to \ell^-\bar{\nu}_\ell X]}{N[\bar{B}^0_d(t)\to \ell^+\nu_\ell X]+N[B_d(t)\to \ell^-\bar{\nu}_\ell X]},
\eea
where the above object is approximated to an excellent precision as
\bea
\mathcal{A}_{\rm SL}^d=\frac{|p/q|^2-|q/p|^2}{|p/q|^2+|q/p|^2}\simeq \mathrm{Im}\left(\frac{\Gamma_{12}}{M_{12}}\right).\label{Eq:CPVio}
\eea
In Eq.~(\ref{Eq:CPVio}), $M_{12}$ and $\Gamma_{12}$ are calculated as complex conjugate of Eqs.~(\ref{Eq:M21}) and (\ref{Eq:Gamma21}).

%====================
\section{Numerical results} \label{Sec:5}
%====================
In the analysis, $\textrm{Re}(\bar{a}_2), \textrm{Im}(\bar{a}_2),$ and $\delta^\prime$ are treated as parameters determined in the numerical result, since they are not predictable within the QCDF approach. As to the color-allowed tree diagram, the coefficient consists of the SM part and NP contributions,
\bea
a_1(m_b)=a_1^{\rm SM}(m_b)+c_1^{\rm NP}(m_b)+\frac{c_2^{\rm NP}(m_b)}{3}.\label{Eq:a1determination}
\eea
For the SM contribution, the universal value of $a_1^{\rm SM}(m_b)=1.070\pm 0.012$ \cite{Endo:2021ifc} is adopted, realized to the high precision \cite{Huber:2016xod} at NNLO. Contributions beyond the SM are included at the scale of $\mu=M_W$,
\bea
c_i^{\rm}(M_W)=c_i^{\rm SM}(M_W)+c_i^{\rm NP}(M_W)\quad
(i=1, 2),\label{Eq:npwilsonco1}
\eea
while the Wilson coefficients of the (chromomagnetic) penguin operators are fixed to the SM values at the same scale. Here $c_i^{\rm NP}(M_W)~(i=1, 2)$ in Eq.~(\ref{Eq:npwilsonco1}) is set to a real-valued parameter and is taken as independent of the flavors, which universally affect $b\to c\bar{q}_2q_3$ for $q_2=u, c$ and $q_3=d, s$. With Eq.~(\ref{Eq:npwilsonco1}), the radiative corrections are discussed separately for the SM and NP, where LO is sufficiently accurate for NP,
\bea
\begin{pmatrix}
	c_1^{\rm NP}(m_b)\\
	c_2^{\rm NP}(m_b)
\end{pmatrix}
=U^{\rm (LO)}
\begin{pmatrix}
	c_1^{\rm NP}(M_W)\\
	c_2^{\rm NP}(M_W)
\end{pmatrix}.
\label{Eq:Wilson12}
\eea
In the above relation, $U^{\rm (LO)}$ can be obtained as it is customarily done \cite{Buchalla:1995vs}.
\par
In what follows, the detail of the numerical investigation is outlined; for definitiveness, one of the six categories in Table~\ref{Tab:1} is discussed, while the other five cases are analyzed in a similar way. We first generate a value of $a_1(m_b)$ randomly from the range
\bea
0<a_1(m_b)<a_1^{\rm max},\label{Eq:random}
\eea
with the upper limit selected to cover the relevant parameter range, $a_1^{\rm max}=1.15$. As a next step, we generate $\mathrm{Br}^{ij}$ with $(i, j)=(+, -), (0, 0), (0, -)$, $V_{cb}, f_T, f_C$ and $f^{B\to D,+-}$ as random Gaussian numbers. Here, $f_T~(f_C)$ represents a decay constant for a meson that is emitted from the $W$ boson in the color-allowed (color-suppressed) tree process while $f^{B\to D, +-}$ represents heavy-to-heavy form factors with proper charge assignment in the final state. The central value and uncertainty for $\mathrm{Br}^{ij}$ are given in Table~\ref{Tab:1} while those for the other ones are given in Table~\ref{Tab:2} of Appendix~\ref{App:B}.\par
With the generated parameters, $\mathrm{Re}(\bar{a}_2), \mathrm{Im}(\bar{a}_2)$ and $\delta^\prime$ are computed from Eqs.~(\ref{Eq:Sol1})-(\ref{Eq:Sol3}) or Eqs.~(\ref{Eq:ReaSU3})-(\ref{Eq:SU3breakingsol}), with 
the choice of overall signs in Eqs.~(\ref{Eq:Sol2}) and (\ref{Eq:ImaSU3}) and
the twofold ambiguity of $\delta^\prime$ in Eqs.~(\ref{Eq:Sol3}) and (\ref{Eq:SU3breakingsol}) selected randomly with a large sampling number.
At this stage, we properly remove the parameter set that does not  satisfy Eqs.~(\ref{Eq:condi1})-(\ref{Eq:condi3}) or Eqs.~(\ref{Eq:a1SU3})-(\ref{Eq:thiSU3}) in such a way to ensure the existence of the solutions.\par
It should be noted that $c_1^{\rm NP}(M_W)$ and $c_2^{\rm NP}(M_W)$ are not simultaneously determined by the given value of $a_1(m_b)$ in Eq.~(\ref{Eq:a1determination}). In view of this aspect, $c_2^{\rm NP}(M_W)$ is computed from the fixed values of $a_1(m_b)$ and $c_1^{\rm NP}(M_W)$, via the relation in Eq.~(\ref{Eq:a1determination}), {\it i.e.,}
\bea
c_2^{\rm NP}(M_W)=\frac{a_1(m_b)-a_1^{\rm SM}(m_b)-(U_{11}^{\rm (LO)}+U_{21}^{\rm (LO)}/3)c_1^{\rm NP}(M_W)}{
	U_{12}^{\rm (LO)}+U_{22}^{\rm (LO)}/3}.\label{Eq:c1formula}
\eea
This means that the possible values of $c_2^{\rm NP}(M_W)$ are scanned in the parameter space. For $a_1^{\rm SM}(m_b)$, its imaginary part arises solely from the radiative correction \cite{Beneke:2000ry, Huber:2016xod} and is negligible to high accuracy for our current purpose.\par
The $\tau(B^+)/\tau(B_d)$ in the presence of NP can be evaluated from $c_2^{\rm NP}(M_W)$ and $c_1^{\rm NP}(M_W)$. In analyzing the lifetime ratio, input parameters including the heavy-quark mass and power correction parameters in the heavy quark effective theory (HQET) are adopted from Ref.~\cite{Bordone:2021oof} in the kinetic scheme \cite{Bigi:1994ga, Bigi:1996si}. As to the value in the SM at NLO QCD, the recent result \cite{Egner:2024lay} is
\bea
\left[\frac{\tau(B^+)}{\tau(B^0_d)}\right]_{\rm SM,\; NLO}=1.081^{+0.014}_{-0.016}\label{Eq:calSMTAU}.
\eea
We adopt the central value and the larger side of the uncertainty in Eq.~(\ref{Eq:calSMTAU}). For the interference terms between SM and NP contributions, as well as the terms purely originating from NP, we consider the LO accuracy in QCD corrections with $c_1^{\rm SM}(m_b)=1.098$ and $c_2^{\rm SM}(m_b)=-0.231$, which can be obtained with Ref.~\cite{Buchalla:1995vs}. The same accuracy is used in the numerical analysis of $B_d-\bar{B}_d$ mixing. For the charm-quark mass input, the $\bar{m}_c(m_c)$ is converted to one at 3~GeV via RunDec \cite{Chetyrkin:2000yt}, leading to $m_c = 0.985~\mathrm{GeV}$.
\par
One can define $\chi^2$ to impose the constraints on the parameters of the NP scenario \cite{Lenz:2022pgw}. In our analysis, the following $\chi^2$ functions are introduced:
\bea
\chi^2_{\rm (A)}&=&\displaystyle\sum_{(i, j)=(+, -)}^{(0, 0), (0, -)}
\left(\frac{\mathrm{Br}^{ij}-\mathrm{Br}^{ij}_{\rm cent}}{\delta \mathrm{Br}^{ij}}\right)^2
+\left(\frac{|V_{cb}|-|V_{cb}|_{\mathrm{cent}}}{\delta |V_{cb}|}\right)^2
+\left(\frac{f_T-f_{T,~\mathrm{cent}}}{\delta f_T}\right)^2
\nn\\
&&
+\left(\frac{f_{C}-f_{C,~\mathrm{cent}}}{\delta f_{C}}\right)^2
+\left(\frac{f^{B\to D, +-}-f^{B\to D, +-}_{\rm cent}}{\delta f^{B\to D, +-}}\right)^2,
\qquad\qquad
\label{Eq:chisquare}\\
\chi^2_{\rm (B)}&=&\chi^2_{\rm (A)}+
\left\{\frac{[\tau(B^+)/\tau(B_d)]_{\rm th}-[\tau(B^+)/\tau(B_d)]_{\rm exp}}
{\sqrt{\delta [\tau(B^+)/\tau(B_d)]^2_{\rm th}
		+\delta [\tau(B^+)/\tau(B_d)]^2_{\rm exp}}}\right\}^2.\qquad\qquad
\label{Eq:chisquareB}
\eea
It should be noted that $\chi^2_{(A)}$ does not include the $\tau(B^+)/\tau(B^+)$ constraint, while $\chi^2_{(B)}$ does. The above two quantities are evaluated based on the parameters generated from the Gaussian distribution as described before. This analysis is not the minimization procedure and instead scans the parameter region \cite{Lenz:2022pgw} in the present case including rescattering for the exclusive decays. In Eq.~(\ref{Eq:chisquare}), $(\cdots)_{\rm cent}$ represents the central value of relevant quantities while $\delta(\cdots)$ stands for its uncertainty given in Tables~\ref{Tab:1} and \ref{Tab:2}. The heavy-to-light form factors are set to their central values and not included in Eqs.~(\ref{Eq:chisquare}) and (\ref{Eq:chisquareB}) since the branching ratios have rather weak dependence on those quantities, which are accompanied by SU(3) breaking, as given in $\Delta_{DP}^{(2)}$ in Eq.~(\ref{Eq:SU3param1}).  As for $|V_{cb}|_{\rm cent}$ and $\delta|V_{cb}|$ in Eq.~(\ref{Eq:chisquare}), we use the value obtained by the exclusive fitting \cite{Iguro:2020cpg} exhibited in Table~\ref{Tab:2}. In Eq.~(\ref{Eq:chisquareB}), the larger theoretical uncertainty of the lifetime ratio in Eq.~(\ref{Eq:calSMTAU}) is adopted as $\delta[\tau(B^+)/\tau(B_d)]_{\rm th}=0.016$. The experimental data from the Heavy Flavor Averaging Group (HFLAV) are set to $[\tau(B^+)/\tau(B_d)]_{\rm exp}=1.078$ and $\delta[\tau(B^+)/\tau(B_d)]_{\rm exp}=0.004$.
\par
Assembling the mentioned procedure, $\chi^2_{(A)}$ and $\chi^2_{(B)}$ can be calculated with d.o.f.~equal to $7$ and $8$, respectively.
The values of $\chi^2_{(A)}\approx 8.18 ~(\chi^2_{(A)}\approx 14.3) $ and 
$\chi^2_{(B)}\approx 9.30~(\chi^2_{(B)}\approx15.8)$ are used to determine the $1\sigma~(2\sigma)$ region that satisfies the phenomenological constraints.
Furthermore, $\Delta \Gamma_{d}$ and $\mathcal{A}^d_{\rm SL}$ are evaluated as resulting predictions satisfying the mentioned constraints.  The explained routine is  repeated with a number of random values for $a_1(m_b)$ in Eq.~(\ref{Eq:random}). Furthermore, different fixed values of $c_1^{\rm NP}(M_W)$ are investigated in the following results.
\par
The input parameters to compute the $B_d^0-\bar{B}_d^0$ mixing are displayed in Table~\ref{Tab:2}. The bottom-quark mass and charm-quark mass are fixed to $\bar{m}_b(m_b)$ and $\bar{m}_c(m_b)$, respectively, while the top-quark mass is set to $\bar{m}_t(m_t)$. In order to get $\bar{m}_c(m_b)$ and $\bar{m}_t(m_t)$, the respective inputs are converted via RunDec \cite{Chetyrkin:2000yt}, giving $\bar{m}_c(m_b)=0.942~\mathrm{GeV}$ and $\bar{m}_t(m_t)=163.3~\mathrm{GeV}$. This procedure is used to compute the contributions induced by NP with the operator basis in Appendix~\ref{App:A2}.\footnote{In Ref.~\cite{Lenz:2006hd} (see also the review in Ref.~\cite{Artuso:2015swg}), the new operator basis was discussed. This leads to a difference in which operator is treated as the leading power one.} As for the SM contribution, we use \cite{Albrecht:2024oyn}
\bea
[\Delta \Gamma_d]_{\rm SM}=(2.7\pm 0.4)\times 10^{-3}~\mathrm{ps}^{-1}, \quad
[\mathcal{A}_{\rm SL}^d]_{\rm SM}=-(5.1\pm 0.5)\times 10^{-4}.\label{Eq:PredDB2}
\eea
For the experimental data of $\Delta \Gamma_d/\Gamma_d$ and $\mathcal{A}_{\rm SL}^d$, the current values are given by HFLAV \cite{HFLAV:2022esi}, 
\bea
\left[\frac{\Delta \Gamma_d}{\Gamma_d}\right]_{\rm HFLAV} &=& 0.001\pm 0.010,\qquad
\left[\mathcal{A}_{\rm SL}^d\right]_{\rm HFLAV}=-0.0021\pm 0.0017.
\eea
For the latter two quantities, the experimental uncertainties are much larger than the theoretical central values in Eq.~(\ref{Eq:PredDB2}). As for the future experimental projection, an improvement of (statistical) uncertainty is expected for $\Delta \Gamma_d/\Gamma_d$ via upgrade II in the LHCb measurement \cite{LHCb:2018roe}. Moreover, the uncertainty of $\mathcal{A}_{\rm SL}^d$ is also reduced due to Runs 1-5 (300$~\mathrm{fb}^{-1}$) data at LHCb \cite{Cerri:2018ypt}. Those future projections read
\bea
\delta \left(\frac{\Delta\Gamma_d}{\Gamma_d}\right)_{\rm future}=1\times 10^{-3},\quad
\delta\left( \mathcal{A}_{\rm SL}^d\right)_{\rm future}=2\times 10^{-4}.
\eea
The above numerics are adopted as the reference values, assuming that the corresponding central values are unchanged from the current HFLAV data.
\par
\begin{figure}[H]
	\centering
	\begin{minipage}[b]{0.32\columnwidth}
		\centering
		\includegraphics[width=\columnwidth]{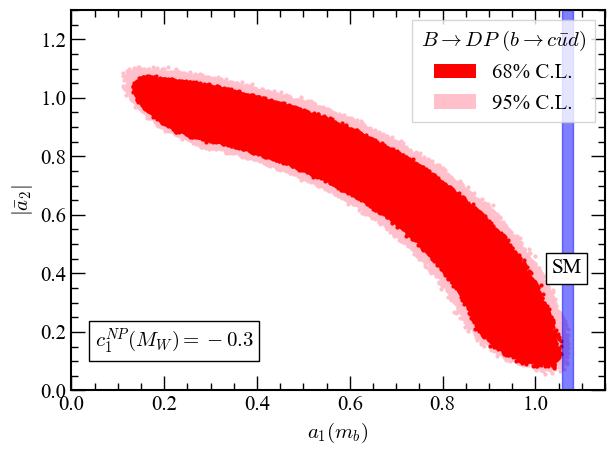}
	\end{minipage}
	\begin{minipage}[b]{0.32\columnwidth}	
		\centering
		\includegraphics[width=\columnwidth]{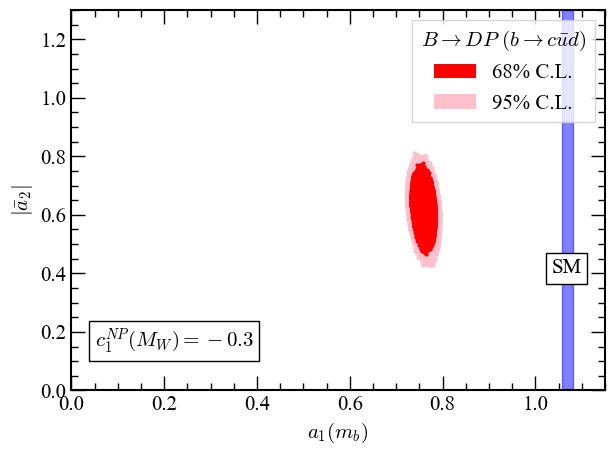}
	\end{minipage}
	\begin{minipage}[b]{0.32\columnwidth}	
	      \centering
		\includegraphics[width=\columnwidth]{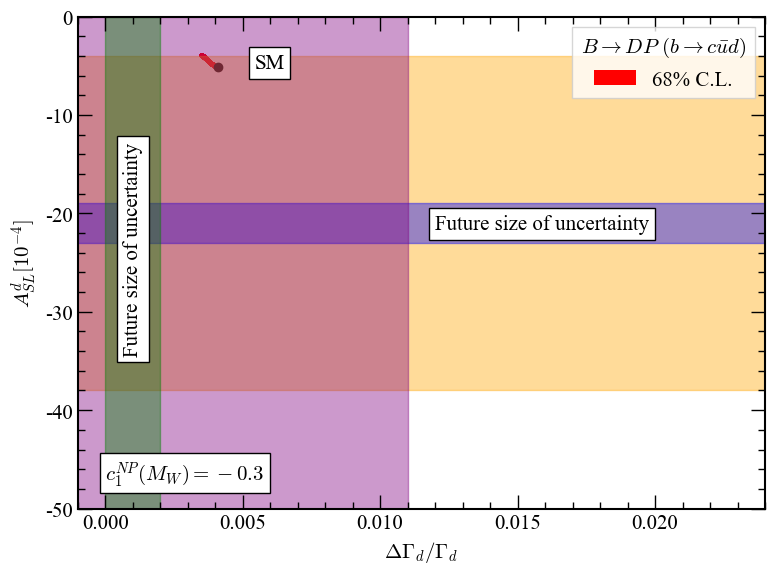}
	\end{minipage}\\
	\begin{minipage}[b]{0.32\columnwidth}
		\centering
		\includegraphics[width=\columnwidth]{Fig1aprime.png}
	\end{minipage}
	\begin{minipage}[b]{0.32\columnwidth}
		\centering
		\includegraphics[width=\columnwidth]{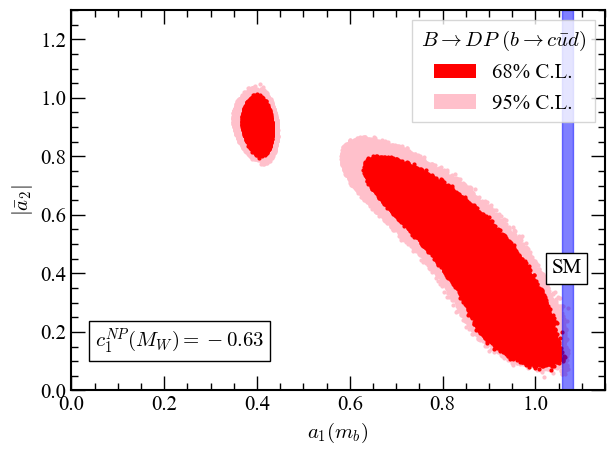}
	\end{minipage}
	\begin{minipage}[b]{0.32\columnwidth}
		\centering
		\includegraphics[width=\columnwidth]{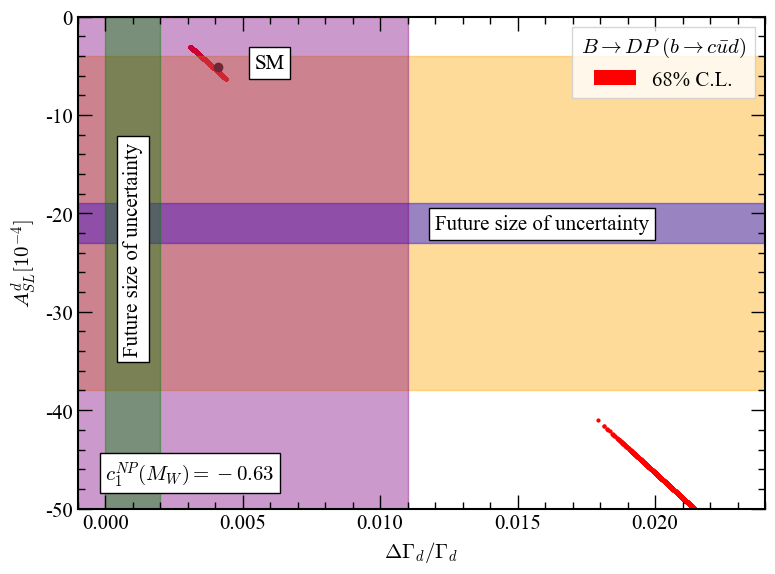}
	\end{minipage}
	\begin{minipage}[b]{0.32\columnwidth}
		\centering
		\includegraphics[width=\columnwidth]{Fig1aprime.png}
	\end{minipage}
	\begin{minipage}[b]{0.32\columnwidth}
		\centering
		\includegraphics[width=\columnwidth]{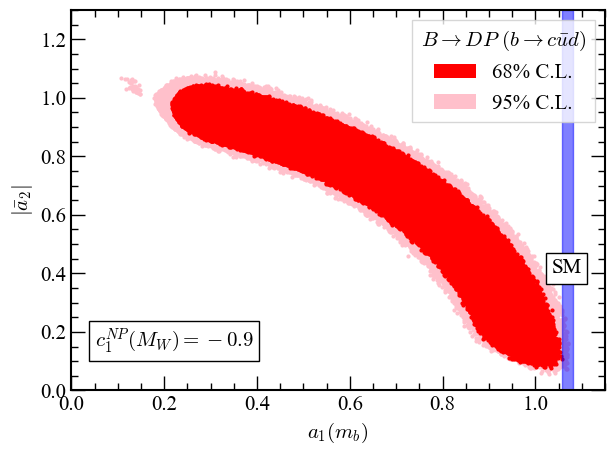}
	\end{minipage}
	\begin{minipage}[b]{0.32\columnwidth}
		\centering
		\includegraphics[width=\columnwidth]{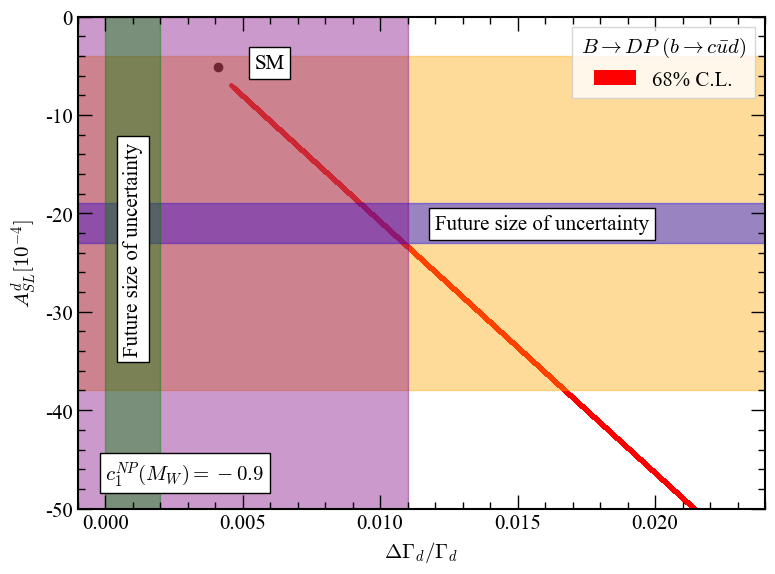}
	\end{minipage}
	\caption{Left column: parameter regions that satisfy the phenomenological constraints  without $\tau(B^+)/\tau(B_d)$, for the $a_1(m_b)$ versus $|\bar{a}_2|$ plane. Middle column: parameter regions that satisfy the constraints including $\tau(B^+)/\tau(B_d)$ in addition to those in left column. See the main text for details. The blue bands represent $a_1^{\rm SM}(m_b)=1.070\pm 0.012$ \cite{Endo:2021ifc}, universal to the high precision \cite{Huber:2016xod} at NNLO. The red and pink points represent the regions where the constraints are satisfied at $1\sigma$ and $2\sigma$ confidence levels, respectively. Right column: predictions for $\Delta\Gamma_d/\Gamma_d$ and $\mathcal{A}_{\rm SL}^d$ that satisfy the phenomenological constraints including the $\tau(B^+)/\tau(B_d)$ data. The central value for the SM prediction is given by a black point while the yellow and light purple bands represent the current HFLAV $1\sigma$ ranges \cite{HFLAV:2022esi}. The future experimental uncertainties \cite{LHCb:2018roe, Cerri:2018ypt}, where the central values are assumed to remain unchanged from those in HFLAV \cite{HFLAV:2022esi}, are represented as purple and green bands. The upper, middle, and lower rows, respectively, represent the results with $c_1^{\rm NP}(M_W)=-0.3, -0.63$ and $-0.9$.}
	\label{Fig:1}
\end{figure}

In order to exhibit how the $\tau(B^+)/\tau(B^+)$ constraint works, we consider three choices of parameters: $c_1^{\rm NP}(M_W)=-0.3, -0.63$ and $-0.9$. For illustrative purposes, we first take $B\to DP$ for the $b\to c\bar{u}d$ transition. In the left column of Fig.~\ref{Fig:1}, the allowed parameter regions that satisfy the phenomenological constraints {\it without} the $\tau(B^+)/\tau(B_d)$ data based on Eq.~(\ref{Eq:chisquare}) are exhibited for the $a_1(m_b)$ versus $|\bar{a}_2|$ plane. These three plots are to be contrasted with those in the middle column in Fig.~\ref{Fig:1}, which account for the $\tau(B^+)/\tau(B_d)$ constraint in addition to those in left column, based on Eq.~(\ref{Eq:chisquareB}). 
The middle column panels give an improved result compared with Ref.~\cite{Endo:2021ifc}, since the constraint of the lifetime ratio is included.
\par
Comparing the left and middle columns of Fig.~\ref{Fig:1}, one immediately finds that how stringent the lifetime constraint is depends crucially on the choice of the NP parameters. Among the displayed results, $c_1^{\rm NP}(M_W)=-0.3$, corresponding to the upper-left and upper-middle panels, gives a result that is most significantly constrained by the lifetime ratio. However, for the case of $c_1^{\rm NP}(M_W)=-0.9$, the lifetime constraint works weakly, as shown in the lower-left and lower-middle panels in Fig.~\ref{Fig:1}.
\par
\begin{figure}[H]
	\centering
	\begin{minipage}[b]{0.49\columnwidth}
		\centering
		\includegraphics[width=0.9\columnwidth]{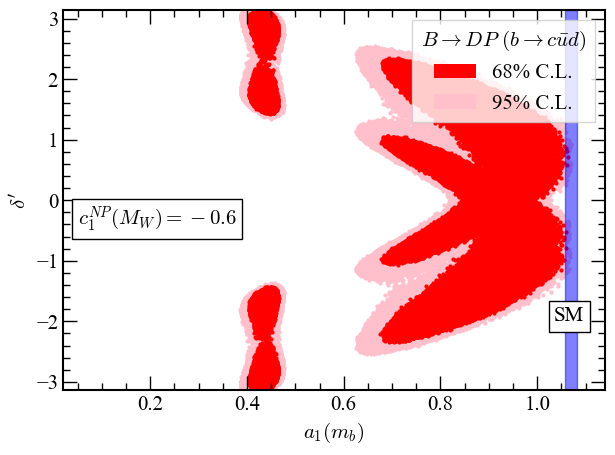}
	\end{minipage}
	\begin{minipage}[b]{0.49\columnwidth}
		\centering
		\includegraphics[width=0.9\columnwidth]{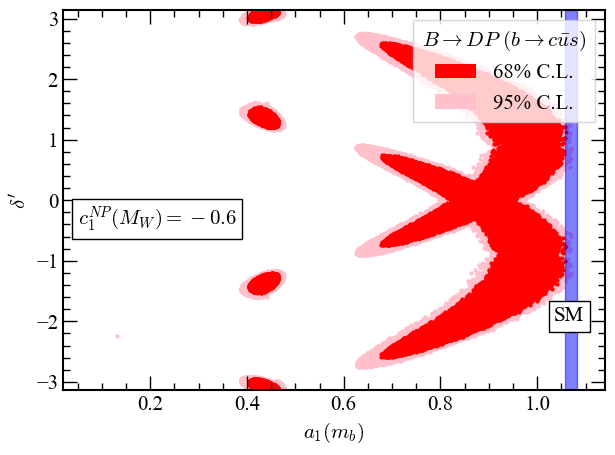}
	\end{minipage}
	\begin{minipage}[b]{0.49\columnwidth}
		\centering
		\includegraphics[width=0.9\columnwidth]{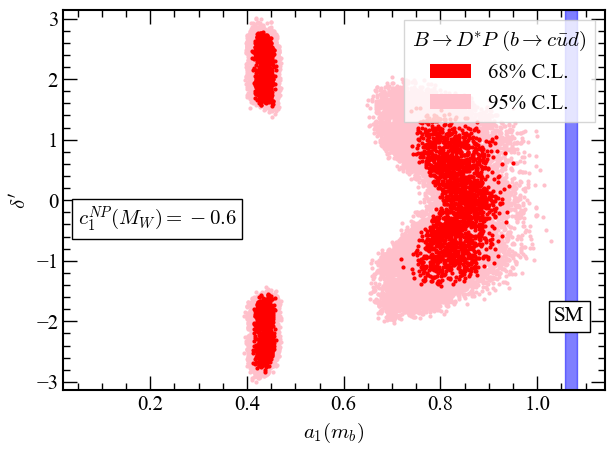}
	\end{minipage}
	\begin{minipage}[b]{0.49\columnwidth}
		\centering
		\includegraphics[width=0.9\columnwidth]{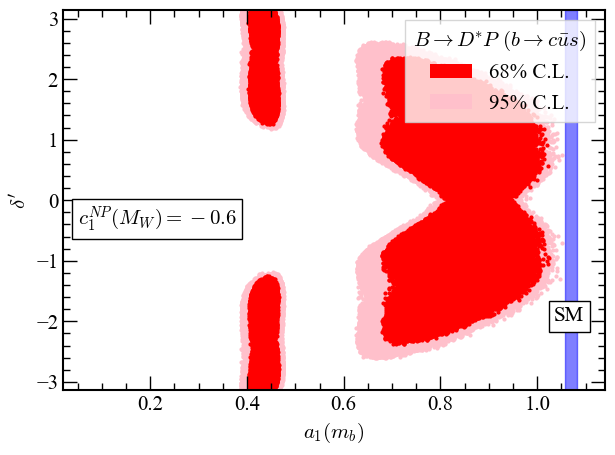}
	\end{minipage}
	\begin{minipage}[b]{0.49\columnwidth}
		\centering
		\includegraphics[width=0.9\columnwidth]{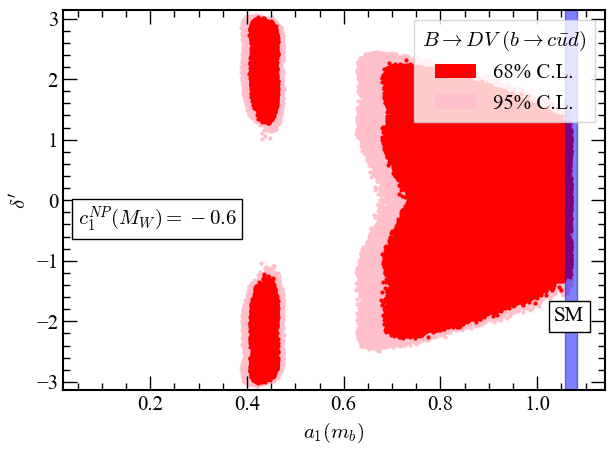}
	\end{minipage}
	\begin{minipage}[b]{0.49\columnwidth}
		\centering
		\includegraphics[width=0.9\columnwidth]{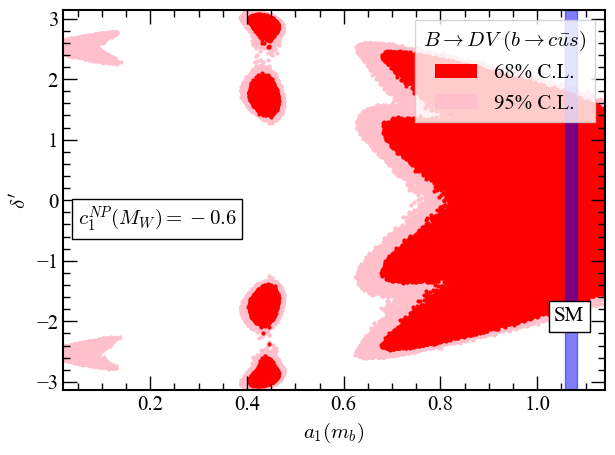}
	\end{minipage}
	\caption{Plots similar to Fig.~2 with the vertical axes replaced by $\delta^\prime$, the rescattering angle. The constraint of the $\tau(B^+)/\tau(B_d)$ data is included in the individual plots.}
	\label{Fig:3}
\end{figure}

Furthermore, in the right column of Fig.~\ref{Fig:1}, the resulting predictions for $B_d^0-\bar{B}_d^0$ mixing are exhibited. The results are based on the parameter region that satisfies the phenomenological constraints including $\tau(B^+)/\tau(B_d)$ for $68\%$ C.L. In order to compute $\Delta \Gamma_d/\Gamma_d$, the formula in Eq.~(\ref{Eq:Ciucalc}) and the HFLAV lifetime of $B_d$ in Eq.~(\ref{Eq:lifetimevalues}) are used. Among the plotted choices of $c_1^{\rm NP}(M_W)$, $-0.3$ gives a prediction that is closest to the SM while the deviation range from the SM becomes wider for $c_1^{\rm NP}(M_W)=-0.63$ and $-0.9$. As can be seen in the middle-right and lower-right panels, the resulting variation ranges are larger than the future size of the experimental uncertainties. Hence, we conclude that this type of scenario, where NP contributions are involved in the presence of rescattering, can be testable via future LHCb measurements.
%============
\begin{figure}[H]
	\centering
	\begin{minipage}[b]{0.49\columnwidth}
		\centering
		\includegraphics[width=0.9\columnwidth]{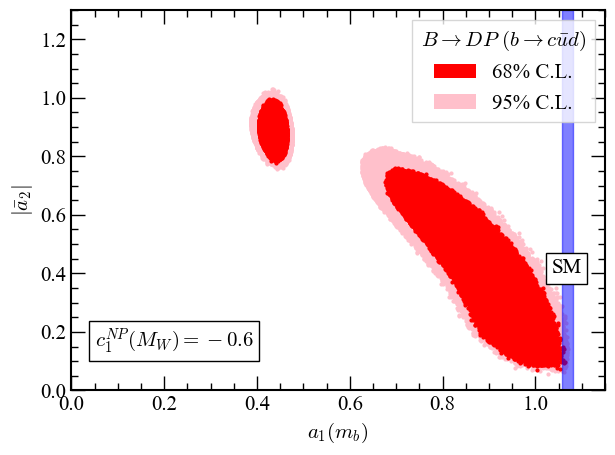}
	\end{minipage}
	\begin{minipage}[b]{0.49\columnwidth}
		\centering
		\includegraphics[width=0.9\columnwidth]{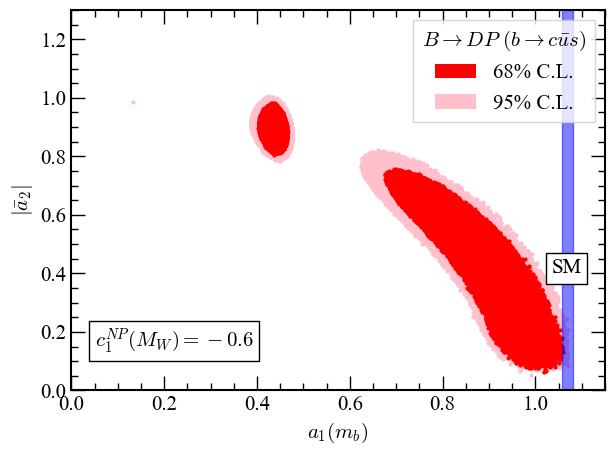}
	\end{minipage}
	\begin{minipage}[b]{0.49\columnwidth}
		\centering
		\includegraphics[width=0.9\columnwidth]{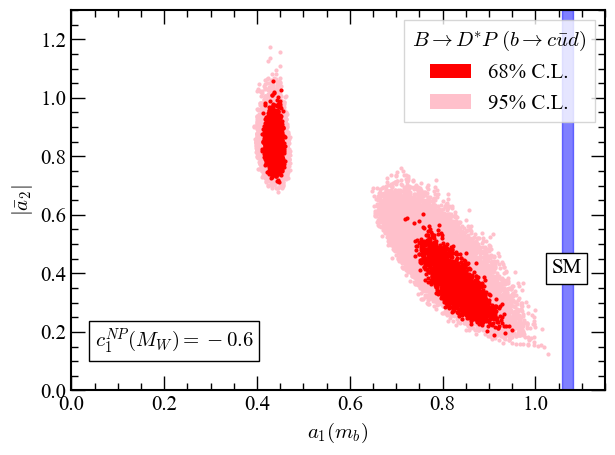}
	\end{minipage}
	\begin{minipage}[b]{0.49\columnwidth}
		\centering
		\includegraphics[width=0.9\columnwidth]{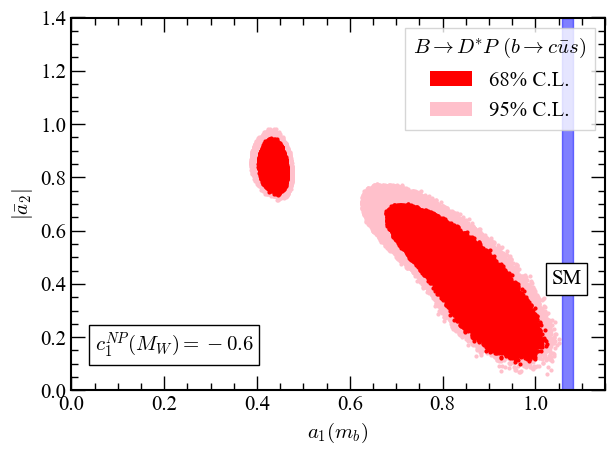}
	\end{minipage}
	\begin{minipage}[b]{0.49\columnwidth}
		\centering
		\includegraphics[width=0.9\columnwidth]{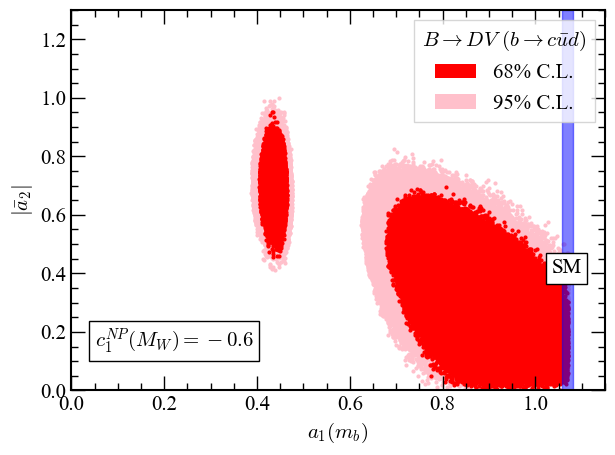}
	\end{minipage}
	\begin{minipage}[b]{0.49\columnwidth}
		\centering
		\includegraphics[width=0.9\columnwidth]{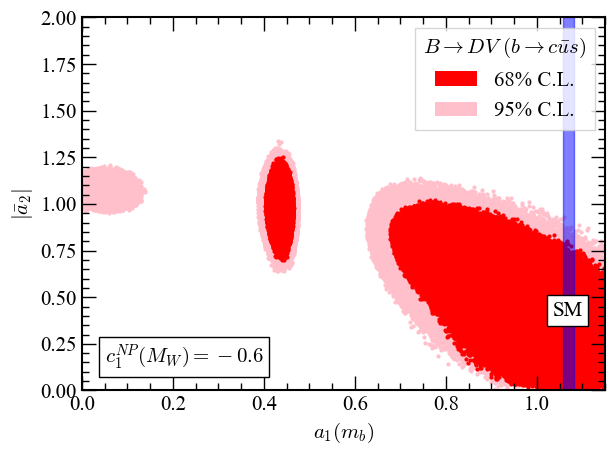}
	\end{minipage}
	\caption{Plots similar to the middle column of Fig.~\ref{Fig:1} except that six different types of final states are analyzed, with fixed $c_1^{\rm NP}(M_W)=-0.6$. The constraint of the $\tau(B^+)/\tau(B_d)$ data is included in the individual plots.}
	\label{Fig:2}
\end{figure}
%============
In Fig.~\ref{Fig:2}, the results similar to the middle column of Fig.~\ref{Fig:1}, except that six different types of final states are analyzed with fixed $c_1^{\rm NP}(M_W)=-0.6$.  As shown in the plots, the patterns of the constrained parameter regions are different individually. Moreover, plots showing the correlation between $a_1(m_b)$ and $\delta^\prime$ are displayed in Fig.~\ref{Fig:3}. It should be noted that the constraint from $\tau(B^+)/\tau(B_d)$ is included in all plots in Figs.~\ref{Fig:2} and \ref{Fig:3}. As can be seen from the plots, the rescattering angle gives a pattern characterized by the sign choice and twofold ambiguity, as explained before.
%============
%============
%=============
\section{Summary}  \label{Sec:6}
%=============
In this work, the phenomenological analysis of $B\to DM$ decays in the presence of quasielastic rescattering was carried out via a model-independent manner, which in general includes the contributions of NP. The rescattering phase and coefficient of the color-suppressed tree diagram (denoted as $a_2^{\rm eff}$) were analytically constrained by the experimental data of the branching ratios and theoretical inputs such as form factors. These feasible restrictions were applied for the final states with $S=-1, I_z=0$ and $S=1, I_z=-1$, where the branching ratios are altered only by the relative phase between $ \delta_{6}$ and $\delta_{\overline{15}}$. The numerical results are given for the two-body nonleptonic decays of $\bar{B}_{(s)}\to D^{(*)}_{(s)}P$ and $\bar{B}_{(s)}\to D_{(s)}V$ in a systematic way. For both $b\to c\bar{u}s$ and $b\to c\bar{u}d$, the set of the constraining relations were obtained, where the latter includes the SU(3) breaking from the decay constants and masses.
\par
We included the $B$-meson lifetime ratio to impose constraints on the phenomenological discussion of $B\to DM$ in the presence of the quasielastic rescattering. These observables are correlated with $B\to DM$ due to the nonleptonic Wilson coefficients. For the NP contributions, we considered the model-independent modification of the Wilson coefficients for the current-current operators, denoted as $c_1$ and $c_2$. Depending on the parameter space, we found that the lifetime ratio can give a stringent bound on the rescattering and NP parameters, as the NP contribution modifies Pauli interference, affecting the lifetime difference between $B^+$ and $B_d$. Meanwhile, it was also found that some specific parameter sets, such as $c_1^{\rm NP}(M_W)=-0.9$ with $c_2^{\rm NP}(M_W)$ varied, are rather weekly constrained by the lifetime ratio. Based on this methodology, the allowed parameter regions for $a_1(m_b)$, $a_2^{\rm eff}$ and $\delta^\prime$ were discussed, where the correlations between them were clarified numerically.
\par
Furthermore, the width difference and \textit{CP} violation in $B_d^0-\bar{B}_d^0$ mixing, where the latter is measured via the semileptonic asymmetry, were analyzed as predictions that satisfy the phenomenological constraints such as $\tau(B^+)/\tau(B_d)$ and $\mathrm{Br}[B\to DM]$. We found that for some specific choices of Wilson coefficients from NP, the two mentioned observables can be considerably shifted from the SM predictions. This deviation size is larger than the future uncertainties in the LHCb experiment \cite{LHCb:2018roe, Cerri:2018ypt}, and thus considered scenario, in which the rescattering and beyond-the-SM contributions are involved, is testable via future measurements.
%=============
\section*{Acknowledgments}
%=============
The authors would like to thank Hai-Yang Cheng for providing the \textit{Mathematica} code. Furthermore, the authors are grateful to Boris Blok and Xiaodong Shi for useful comments. The work of H.~U.~is supported by the National Science Foundation of China under Grant No.~12405111 and the Seeds Funding of Jilin University.
%=============
\section*{Data Availabity}
%=============
No data were created or analyzed in this study.
\let\origclearpage\clearpage
\let\clearpage\relax
\begin{appendices} \renewcommand{\thesection}{\Alph{section}}
	\counterwithin*{equation}{section}
	\renewcommand\theequation{\thesection\arabic{equation}}
      \section{Determinations of $\bar{a}_2$ and $\delta^\prime$ from experimental data}\label{App:0}
\subsection{$b\to c \bar{u}s$}\label{App:01}
Here, the derivations of Eqs.~(\ref{Eq:Sol1}), (\ref{Eq:Sol2}), and (\ref{Eq:Sol3}) are given. The coefficients in Eq.~(\ref{Eq:TDK}) are defined by
\bea
N^T_{D K}&=&\frac{G_F}{\sqrt{2}}(m_{B}^2-m_{D}^2)f_{K}F_0^{BD}(m_{K}^2),\quad
N^C_{D K}=\frac{G_F}{\sqrt{2}}(m_{B}^2-m_{K}^2)f_{D}F_0^{BK}(m_{D}^2),\qquad\quad\label{Eq:Nfactors}\\
N^T_{D^*K}&=&\frac{G_F}{\sqrt{2}}2m_{D^*}f_KA_0^{BD^*}(m_K^2),\quad\quad\;\;\:
N^C_{D^*K}=\frac{G_F}{\sqrt{2}}2m_{D^*}f_{D^*}F_+^{BK}(m_{D^*}^2),\label{Eq:Nfactors2}\\
N^T_{DK^*}&=&\frac{G_F}{\sqrt{2}}2m_{K^*}f_{K^*}F_+^{BD}(m_{K^*}^2),\quad\quad\;\:
N^C_{DK^*}=\frac{G_F}{\sqrt{2}}2m_{K^*}f_{D}A_0^{BK^*}(m_{D}^2).\label{Eq:Nfactors3}
\eea
In what follows, we consider $B$-meson decays into two pseudoscalars for definitiveness unless otherwise specified. In the presence of quasielastic rescattering, amplitudes are given by
\bea
\mathcal{A}_f^{+-}&=&N^T_{DK}
a_1\left(\frac{1+e^{i\delta^\prime}}{2}+\bar{a}_2\frac{1-e^{i\delta^\prime}}{2}\right)e^{i\delta_{\overline{15}}},\label{Eq:a}\\
\mathcal{A}_f^{00}&=&N^T_{DK}a_1\left(\frac{1-e^{i\delta^\prime}}{2}+\bar{a}_2\frac{1+e^{i\delta^\prime}}{2}\right)e^{i\delta_{\overline{15}}},\label{Eq:b}\\
\mathcal{A}_f^{0-}&=& N^T_{DK}a_1\left(1+\bar{a}_2\right).\label{Eq:c}
\eea
One can find that dependence on the heavy-to-light form factors is absorbed by $\bar{a}_2$, so that Eqs.~(\ref{Eq:a})-(\ref{Eq:c}) can be evaluated solely by the heavy-to-heavy form factors. This is not the case for $b\to c\bar{u}d$ decays, as explicitly shown later.\par
It should be noted that the overall phase in Eqs.~(\ref{Eq:a})-(\ref{Eq:c}) cancels out when being squared for the evaluation of decay rates. By substituting Eqs.~(\ref{Eq:a})-(\ref{Eq:c}) into Eq.~(\ref{Eq:branch1}), one can obtain the branching ratios
\bea
\frac{\mathrm{Br}^{+-}}{\mathcal{N}_{DK}}&=&\frac{1+\cos\delta^\prime}{2}+|\bar{a}_2|^2\frac{1-\cos\delta^\prime}{2}+\mathrm{Im}(\bar{a}_2)\sin\delta^\prime,\label{Eq:d}\\
\frac{\mathrm{Br}^{00}}{\mathcal{N}_{DK}}&=&\frac{1-\cos\delta^\prime}{2}+|\bar{a}_2|^2\frac{1+\cos\delta^\prime}{2}-\mathrm{Im}(\bar{a}_2)\sin\delta^\prime,\qquad\quad
\label{Eq:e}\\
\frac{\tau^{+-}}{\tau^{0-}}\frac{\mathrm{Br}^{0-}}{\mathcal{N}_{DK}}&=&
1+|\bar{a}_2|^2+2\mathrm{Re}(\bar{a}_2),\label{Eq:f}
\eea
where the following objects are introduced:
\bea
\mathcal{N}_{DK}&=&\frac{\tau_{P}p_{\rm cm}[P\to M_1M_2]}{8\pi m_P^2}|V_{cb}V_{us}^*|^2(N^T_{M_1M_2})^2|a_1|^2,\label{Eq:carN}\\
\mathcal{N}_{D^*K}&=&\frac{\tau_{P}p_{\rm cm}^3[P\to M_1^*M_2]}{8\pi m_{M_1^*}^2}|V_{cb}V_{us}^*|^2(N^T_{M_1^*M_2})^2|a_1|^2,\\
\mathcal{N}_{DK^*}&=&\frac{\tau_{P}p_{\rm cm}^3[P\to M_1M_2^*]}{8\pi m_{M_2^*}^2}|V_{cb}V_{us}^*|^2(N^T_{M_1M_2^*})^2|a_1|^2.
\eea
Furthermore, the following variables are introduced:
\bea
A_{DK}&=&2\mathrm{Im}(\bar{a}_2),\\
B_{DK}&=&1-|\bar{a}_2|^2,\\
\omega_{DK} &=&\begin{cases}
	\mathrm{Arcsin}\left(\frac{B_{DK}}{\sqrt{A^2_{DK}+B^2_{DK}}}\right) & \mathrm{for}~A_{DK}\geq 0,\\
	\pi\mathrm{sign}(B_{DK})-\mathrm{Arcsin}\left(\frac{B_{DK}}{\sqrt{A_{DK}^2+B_{DK}^2}}\right) & \mathrm{for}~A_{DK}< 0.
\end{cases}\label{Eq:omegaDK}
\eea
By rewriting the three relations in Eqs.~(\ref{Eq:d}), (\ref{Eq:e}), and (\ref{Eq:f}) in terms of $\textrm{Re}(\bar{a}_2), \textrm{Im}(\bar{a}_2)$ and $\delta^\prime$, one can obtain Eqs.~(\ref{Eq:Sol1})-(\ref{Eq:Sol3}) if the conditions of Eqs.~(\ref{Eq:condi1})-(\ref{Eq:condi3}) are satisfied. 
%=====
\subsection{$b\to c\bar{u}d$}\label{App:02}
%=====
The derivation of Eqs.~(\ref{Eq:ReaSU3})-(\ref{Eq:SU3breakingsol}) is given in a way similar to $b\to c\bar{u}s$ decays except that the SU(3) breaking should be taken into account. We introduce parameters related to SU(3) breaking,
\bea
z_{DP}&=&\frac{f_{D_s}f_\pi}{f_D f_K},\qquad\:\:\:
r_{DP}=\frac{p_{\rm cm}[\bar{B}^0_s\to D^0\bar{K}^0]}{p_{\rm cm}[\bar{B}^0_s\to D_s^+\pi^-]},\\
z_{D^*P}&=&\frac{f_{D_s^*}f_\pi}{f_{D^*} f_K},\qquad
r_{D^*P}=\left(\frac{m_{D_s^{*+}}}{m_{D^{*0}}}\right)^2\frac{p_{\rm cm}^3[\bar{B}^0_s\to D^{*0}\bar{K}^0]}{p_{\rm cm}^3[\bar{B}^0_s\to D_s^{*+}\pi^-]},\\
z_{DV}&=&\frac{f_{D_s}f_\rho}{f_{D} f_{K^*}},\qquad\:\:
r_{DV}=\left(\frac{m_{\rho^{+}}}{m_{K^{*0}}}\right)^2\frac{p_{\rm cm}^3[\bar{B}^0_s\to D^0\bar{K}^{*0}]}{p_{\rm cm}^3[\bar{B}^0_s\to D_s^+\rho^-]},
\eea
\vspace{-7mm}
\bea
\Delta^{(1)}_{DP}&=&\left(\frac{N_T^{D^+_s\pi^-}}{N_T^{D^{0}\pi^-}}\frac{N_C^{D^{0}\pi^-}}{N_C^{D^0\bar{K}^0}}\right)^{-1}\frac{\mathcal{N}_{D_s^+\pi^-}}{\mathcal{N}_{D^0\pi^-}}\frac{\tau^{0-}}{\tau^{+-}}
-1,\qquad
\Delta^{(2)}_{DP}=
\frac{1}{z_{DP}^2}\frac{N_T^{D^+_s\pi^-}}{N_T^{D^{0}\pi^-}}\frac{N_C^{D^{0}\pi^-}}{N_C^{D^0\bar{K}^0}}-1,
\label{Eq:SU3param1}\qquad\qquad\\
\Delta^{(3)}_{DP}&=&\frac{z_{DP}^2}{r_{DP}}-1,
\quad \Delta^{(4)}_{DP}=z_{DP}^2\left(\frac{N_T^{D^+_s\pi^-}}{N_T^{D^{0}\pi^-}}\frac{N_C^{D^{0}\pi^-}}{N_C^{D^0\bar{K}^0}}\right)^{-2}-1,
\quad\Delta^{(5)}_{DP}=\frac{1}{z_{DP}^2}-1.\label{Eq:SU3param2}
\eea

On the basis of the previously introduced notations, the decay amplitudes for $b\to c\bar{u}d$ processes with FSIs can be given as follows:
\bea
\mathcal{A}_f[\bar{B}^0_s\to D^+_s\pi^-]&=&N^T_{D^+_s\pi^-}a_1\left(\frac{1+e^{i\delta^\prime}}{2}+z_{DP}\bar{a}_2\frac{1-e^{i\delta^\prime}}{2}\right)e^{i\delta_{\overline{15}}},\label{Eq:SU3brecase1}\\
\mathcal{A}_f[\bar{B}^0_s\to D^0\bar{K}^0]
&=&\frac{N^T_{D^+_s\pi^-}}{z_{DP}}a_1\left(\frac{1-e^{i\delta^\prime}}{2}+z_{DP}\bar{a}_2\frac{1+e^{i\delta^\prime}}{2}\right)e^{i\delta_{\overline{15}}},
\label{Eq:SU3brecase2}\\
\mathcal{A}_f[B^-\to D^0\pi^-]&=&N^T_{D^0\pi^-}a_1
\left(1+\frac{N^T_{D^+_s\pi^-}}{N^T_{D^0\pi^-}}
\frac{N^C_{D^0\pi^-}}{N^C_{D^0\bar{\pi}^0}}\bar{a}_2\right).\label{Eq:SU3brecase3}
\eea
Since $\bar{a}_2$ is defined so as to absorb $N_C^{D^0\bar{K}^0}$, the overall dependence on heavy-to-light form factors vanishes in Eqs.~(\ref{Eq:SU3brecase1}) and (\ref{Eq:SU3brecase2}), whereas it is included as a prefactor of $\bar{a}_2$ in Eq.~(\ref{Eq:SU3brecase3}). Furthermore, the following parameters are introduced:
\bea
A_{DP}&=&2z_{DP}\mathrm{Im}(\bar{a}_2),\\
B_{DP}&=&1-z_{DP}^2|\bar{a}_2|^2.
\eea
The expression of $\omega_{DP}$ is found by the replacement of $A_{DK}\to A_{DP}$ and $B_{DK}\to B_{DP}$ for $\omega_{DP}$ in Eq.~(\ref{Eq:omegaDK}).
\par
By using the SU(3)-breaking parameters, one can write relations similar to
Eqs.~(\ref{Eq:d})-(\ref{Eq:f}) in the case of $b\to c\bar{u}d$ decays, which is omitted here. These relations are solved with respect to the QCDF approach and the rescattering parameters, leading to Eqs.~(\ref{Eq:ReaSU3})-(\ref{Eq:SU3breakingsol}) under the conditions of Eqs.~(\ref{Eq:a1SU3})-(\ref{Eq:thiSU3}).

	\section{Effective weak operators and matrix elements}\label{App:A}
	\subsection{$\Delta B=1$ processes}\label{App:A1}
	The effective operators for the weak Hamiltonian in Eq.~(\ref{Eq:BurasHw}) are defined by \cite{Lenz:2022pgw},
	\bea
	Q_1^{\bar{q}_2q_3}&=&(\bar{c}^\alpha b^\alpha)_{\rm V-A}(\bar{q}_3^\beta q_2^\beta)_{\rm V-A},\qquad\qquad\hspace{3mm}
	Q_2^{\bar{q}_2q_3}=(\bar{c}^\alpha  b^\beta)_{\rm V-A}(\bar{q}_3^\beta q_2^\alpha)_{\rm V-A},\\
	Q_3^{q_3}&=&(\bar{q}_{3}^\alpha b^\alpha)_{\rm V-A}(\bar{q}^\beta q^\beta)_{\rm V-A},\qquad\qquad\quad\;\:
	Q_4^{q_3}=(\bar{q}_{3}^\alpha b^\beta)_{\rm V-A}(\bar{q}^\beta q^\alpha)_{\rm V-A},\\
	Q_5^{q_3}&=&(\bar{q}_{3}^\alpha b^\alpha)_{\rm V-A}(\bar{q}^\beta q^\beta)_{\rm V+A},\qquad\qquad\quad\;\:
	Q_6^{q_3}=(\bar{q}_{3}^\alpha b^\beta)_{\rm V-A}(\bar{q}^\beta q^\alpha)_{\rm V+A},\\
	Q_8^{q_3}&=&\frac{g_s}{8\pi^2}m_b\bar{q}_{3}^\alpha \sigma^{\mu\nu}(1+\gamma_5)t^a_{\alpha\beta}b^\beta G_{\mu\nu}^{a},
	\eea
	where sums over colors denoted by $\alpha$ and $\beta$ and flavor indices are taken implicitly. For $(\cdots)_{\mathrm{V}\pm \mathrm{A}}$, the current is represented as $\gamma^\mu(1\pm \gamma_5)$.
	\par
	As for $B$-meson decays into an exclusive hadronic state, matrix elements relevant for our work are parametrized by form factors \cite{Beneke:2000ry, Neubert:1993mb},
	\bea
	\bra{P(p^\prime)}c\gamma^\mu b\ket{B(p)}&=&F_+^{BP}(q^2)\left[(p+p^\prime)^\mu-
	\frac{m_B^2-m_D^2}{q^2}q^\mu\right]+F_0^{BP}(q^2)\frac{m_B^2-m_D^2}{q^2}q^\mu,\quad\qquad\\
	\bra{V(p^\prime, \epsilon)}c\gamma^\mu\gamma_5 b\ket{B(p)}&=&
	\left[(m_B+m_V)\epsilon^{*\mu}A_1^{BV}(q^2)-\frac{\epsilon^*\cdot q}{m_B+m_V}(p+p^\prime)^\mu A_2^{BV}(q^2)\right.\nn\\
	&&\left.-2m_V\frac{\epsilon^*\cdot q}{q^2}q^\mu A_3^{BV}(q^2)\right]
	+2m_{V}\frac{\epsilon^*\cdot q}{q^2}q^\mu A_0^{BV}(q^2),\\
	A_3^{BV}(q^2)&=&\frac{m_B+m_V}{2m_V}A_1^{BV}(q^2)-\frac{m_B-m_V}{2m_V}A_2^{BV}(q^2),
	\eea
	where $P$ and $V$ are a pseudoscalar and  vector meson, respectively, with $q=p-p^\prime$.
	%=============
	\subsection{$\Delta B=0$ processes}\label{App:A2}
	%=============
	Operators for the $\Delta B=0$ transition are divided into two-quark and four-quark operators. For the former, the dimension-five operators are defined by \cite{Lenz:2022rbq}
	\bea
	O_{\pi}&=&-\bar{b}_v(iD_\mu)(iD^\mu)b_v,\\
	O_{G}&=&\bar{b}_v(iD_\mu)(iD_\nu)(-i\sigma^{\mu\nu})b_v,
	\eea
	where $b(x)=e^{-im_b v\cdot x}b_v(x)$. The matrix elements for the above operators are
	\bea
	\mu_\pi^2 =\frac{\braket{B|O_{\pi}|B}}{2m_{B}},\quad
	\mu_G^2 =\frac{\braket{B|O_G|B}}{2m_{B}}.\label{Eq:HQETMATELE}
	\eea
The matrix elements in Eq,~(\ref{Eq:HQETMATELE}) enter our analysis in the denominator of the second term in Eq.~(\ref{Eq:lifetimeratio}).
	As for the four-quark operators, we introduce \cite{Cheng:2018rkz}
	\bea
	Q_1^q&=&(\bar{b}q)_{\mathrm{V}-\mathrm{A}}(\bar{q}b)_{\mathrm{V}-\mathrm{A}},\label{Eq:opeQ1}\\
	Q_2^q&=&(\bar{b}q)_{\mathrm{S}-\mathrm{P}}
	(\bar{q}b)_{\mathrm{S}+\mathrm{P}},\\
	Q_3^q&=&(\bar{b}t^aq)_{\mathrm{V}-\mathrm{A}}(\bar{q}t^ab)_{\mathrm{V}-\mathrm{A}},\\
	Q_4^q&=&(\bar{b}t^aq)_{\mathrm{S}-\mathrm{P}}
	(\bar{q}t^ab)_{\mathrm{S}+\mathrm{P}},\label{Eq:opeQ4}
	\eea
	where $(\cdots)_{\mathrm{S}\pm \mathrm{P}}$ represents a bilinear of the form $(1\pm \gamma_5)$. The matrix elements for Eqs.~(\ref{Eq:opeQ1})-(\ref{Eq:opeQ4}) are defined by
	\bea
	\braket{B_q|Q_1^q|B_q}&=&f_{B_q}^2m_{B_q}^2B_1,\label{Eq:Matele4quark1}\\
	\braket{B_q|Q_2^q|B_q}&=&f_{B_q}^2m_{B_q}^2B_2,\\
	\braket{B_q|Q_3^q|B_q}&=&f_{B_q}^2m_{B_q}^2\epsilon_1,\\
	\braket{B_q|Q_4^q|B_q}&=&f_{B_q}^2m_{B_q}^2\epsilon_2.\label{Eq:Matele4quark4}
	\eea
	%====================
	\subsection{$\Delta B=2$ processes}\label{App:A3}
	%====================
	Effective operators relevant for $B_d^0-\bar{B}_d^0$ mixing are given by dimension-six operators,
	\bea
	\mathcal{O}_1^d&=&(\bar{b}^\alpha d^\alpha)_{\rm V-A}(\bar{b}^\beta d^\beta)_{\rm V-A},\
	\mathcal{O}_2^d=(\bar{b}^\alpha d^\alpha)_{\rm S-P}(\bar{b}^\beta d^\beta)_{\rm S-P},\\
	\mathcal{O}_3^d&=&(\bar{b}^\alpha d^\beta)_{\rm S-P}(\bar{b}^\beta d^\alpha)_{\rm S-P},\:\;\;
	\mathcal{O}_4^d=(\bar{b}^\alpha d^\alpha)_{\rm S-P}(\bar{b}^\beta d^\beta)_{\rm S+P},\\
	\mathcal{O}_5^d&=&(\bar{b}^\alpha d^\beta)_{\rm S-P}(\bar{b}^\beta d^\alpha)_{\rm S+P},
	\eea
	as well as those giving $1/m_b$ suppressed contributions \cite{Beneke:1996gn, Ciuchini:2003ww},
	\bea
	R_1^d&=&\frac{m_d}{m_b}(\bar{b}^\alpha d^\alpha)_{\rm S-P}(\bar{b}^\beta d^\beta)_{\rm S+P},\\
	R_2^d&=&\frac{1}{m_b^2}[\bar{b}^\alpha \overleftarrow{D}_\rho \gamma^\mu(1-\gamma_5)D^\rho q^\alpha][\bar{b}^\beta \gamma_\mu(1-\gamma_5)q^\beta],\\
	R_3^d&=&\frac{1}{m_b^2}[\bar{b}^\alpha \overleftarrow{D}_\rho (1-\gamma_5)D^\rho q^\alpha][\bar{b}^\beta (1-\gamma_5)q^\beta],\\
	R_4^d&=&\frac{1}{m_b}[\bar{b}^\alpha (1-\gamma_5)iD_\mu q^\alpha][\bar{b}^\beta \gamma^\mu(1-\gamma_5)q^\beta].
	\eea
	The matrix element of the operators are given by
	\bea
	\bra{\bar{B}_d}\mathcal{O}_1^d\ket{B_d}&=&\frac{8}{3}f_{B_d}^2m_{B_d}^2B_1^d,\qquad\quad\qquad\qquad\;
	\bra{\bar{B}_d}\mathcal{O}_2^d\ket{B_d}=-\frac{5}{3}f_{B_d}^2m_{B_d}^2\left(\frac{m_{B_d}}{m_b+m_d}\right)^2B_2^d,\\
	\bra{\bar{B}_d}\mathcal{O}_3^d\ket{B_d}&=&\frac{1}{3}f_{B_d}^2m_{B_d}^2\left(\frac{m_{B_d}}{m_b+m_d}\right)^2B_3^d,\quad
	\bra{\bar{B}_d}\mathcal{O}_4^d\ket{B_d}=2f_{B_d}^2m_{B_d}^2\left(\frac{m_{B_d}}{m_b+m_d}\right)^2B_4^d,\\
	\bra{\bar{B}_d}\mathcal{O}_5^d\ket{B_d}&=&\frac{2}{3}f_{B_d}^2m_{B_d}^2\left(\frac{m_{B_d}}{m_b+m_d}\right)^2B_5^d,\\
	\bra{\bar{B}_d}R_1^d\ket{B_d}&=&\frac{7}{3}\frac{m_d}{m_b}f_{B_d}^2m_{B_d}^2B_{R_1}^d,\quad\qquad\qquad\:
	\bra{\bar{B}_d}R_2^d\ket{B_d}=-\frac{2}{3}f_{B_d}^2m_{B_d}^2\left(\frac{m_{B_d}^2}{m_b^2}-1\right)B_{R_2}^d,\\
	\bra{\bar{B}_d}R_3^d\ket{B_d}&=&\frac{7}{6}f_{B_d}^2m_{B_d}^2\left(\frac{m_{B_d}^2}{m_b^2}-1\right)B_{R_3}^d,\quad
	\bra{\bar{B}_d}R_4^d\ket{B_d}=-f_{B_d}^2m_{B_d}^2\left(\frac{m_{B_d}^2}{m_b^2}-1\right)B_{R_4}^d.\qquad\qquad
	\eea
	It should be noted that the matrix element of $R_1^d$ vanishes in the massless limit of down quark. As for $R_4^d$, the operator is related to other ones \cite{Ciuchini:2003ww},
	\bea
	R_4^q=\frac{1}{4}\mathcal{O}_1^q+\frac{1}{2}\mathcal{O}_2^q+\frac{1}{2}\mathcal{O}_3^q-\frac{m_q}{m_b}\mathcal{O}_5^q+\frac{1}{2}R_2^q.
	\eea
	Hence, $\bra{\bar{B}_d}R_4^d\ket{B_d}$ can be represented by other matrix elements, which is used in our numerical result.
	\section{Numerical input}\label{App:B}
	The experimental values of branching ratios for $B\to DM$ decays are extracted from the 2024 Particle Data Group publication \cite{ParticleDataGroup:2024cfk} and given in Table~\ref{Tab:1}.
	\begin{table}[H]
		\caption{Experimental data of branching ratios of $B$-meson nonleptonic decays. One for $B^-\to D^0\rho^-$ is from Belle II \cite{Belle-II:2024vfw}, while the others are extracted from \cite{ParticleDataGroup:2024cfk}.}
		\label{Tab:1}
		\centering
		\begin{tabular}{|cc|cc|}\hline
			\multicolumn{2}{|c|}{$B\to DP\quad(b\to c\bar{u}d)$}&
			\multicolumn{2}{|c|}{$B\to DP\quad(b\to c\bar{u}s)$}\\\hline
			$B^0_s\to D_s^-\pi^+$&$(2.98\pm0.14)\times 10^{-3}$&
			$B^0\to D^-K^+$&$(2.05\pm 0.08)\times 10^{-4}$\\
			$B^0_s\to \bar{D}^0\bar{K}^0$&$(4.3\pm0.9)\times 10^{-4}$&
			$B^0\to \bar{D}^0K^0$&$(5.5\pm 0.4)\times 10^{-5}$\\
			$B^+\to \bar{D}^0\pi^+$&$(4.61\pm0.10)\times 10^{-3}$&
			$B^+\to \bar{D}^0K^+$&$(3.64\pm 0.15)\times 10^{-4}$\\\hline
			\multicolumn{2}{|c|}{$B\to D^*P\quad(b\to c\bar{u}d)$}&
			\multicolumn{2}{|c|}{$B\to D^*P\quad(b\to c\bar{u}s)$}\\\hline
			$B^0_s\to D_s^{*-}\pi^+$&$(1.9^{+0.5}_{-0.4})\times 10^{-3}$&
			$B^0\to D^{*-}K^+$&$(2.16\pm 0.08)\times 10^{-4}$\\
			$B^0_s\to \bar{D}^{*0}\bar{K}^0$&$(2.8\pm1.1)\times 10^{-4}$&
			$B^0\to \bar{D}^{*0}K^0$&$(3.6\pm 1.2)\times 10^{-5}$\\
			$B^+\to \bar{D}^{*0}\pi^+$&$(5.17\pm0.15)\times 10^{-3}$&
			$B^+\to \bar{D}^{*0}K^+$&$(4.19^{+0.31}_{-0.28})\times 10^{-4}$\\\hline
			\multicolumn{2}{|c|}{$B\to DV\quad(b\to c\bar{u}d)$}&
			\multicolumn{2}{|c|}{$B\to DV\quad(b\to c\bar{u}s)$}\\\hline
			$B^0_s\to D_s^-\rho^{+}$&$(6.8\pm1.4)\times 10^{-3}$&
			$B^0\to D^-K^{*+}$&$(4.5\pm 0.7)\times 10^{-4}$\\
			$B^0_s\to \bar{D}^0\bar{K}^{*0}$&$(4.4\pm0.6)\times 10^{-4}$&
			$B^0\to \bar{D}^0K^{*0}$&$(4.5\pm 0.6)\times 10^{-5}$\\
			$B^-\to D^0\rho^-$&$(9.39\pm 0.21\pm 0.50)\times 10^{-3}$&
			$B^+\to \bar{D}^0K^{*+}$&$(5.3\pm 0.4)\times 10^{-4}$\\\hline
		\end{tabular}
	\end{table}
	The experimental values of the $B$-meson lifetimes from HFLAV \cite{HFLAV:2022esi} are given by
	\bea
	\tau(B^+)=(1.638\pm 0.004)~\mathrm{ps},\quad
	\tau(B_d)=(1.519\pm 0.004)~\mathrm{ps},\quad
	\tau(B_s)=(1.520\pm 0.005)~\mathrm{ps}.\qquad\quad\label{Eq:lifetimevalues}
	\eea
Other input parameters necessary to implement the analysis are given in Table~\ref{Tab:2}.\clearpage
	\begin{center}
		\begin{table}[H]
			\caption{Input parameters given in units of proper powers of GeV. For the parameters in $\Delta B=0$ processes \cite{Bordone:2021oof}, $m_b^{\rm kin}$, $\mu_\pi^2$ and $\mu_G^2$ are defined via the kinetic scheme \cite{Bigi:1994ga, Bigi:1996si} with the hard Wilsonian cutoff at $1~\mathrm{GeV}$. The bag parameters for dimension-six operators relevant to $\Delta B=2$ processes \cite{DiLuzio:2019jyq} are based on the weighted average of the HQET sum rules and lattice QCD. For the form factors, the numerics in Table~4 of Ref.~\cite{Endo:2021ifc} are adopted, which are based on the recent phenomenological fit in Ref.~\cite{Iguro:2020cpg} for the heavy-to-heavy form factors and on Refs.~\cite{Bharucha:2015bzk, Ball:2004ye, Khodjamirian:2017fxg} for the heavy-to-light form factors.}
			\label{Tab:2}
			\centering
			\begin{tabular}{|ccc|ccc|}\hline
				$\alpha_s(M_Z)$ & $0.1180\pm 0.0009$ &\cite{ParticleDataGroup:2024cfk}
				&
				$M_W$ & $80.3692\pm 0.0133$ & \cite{ParticleDataGroup:2024cfk}
				\\
				$\sin\theta_{12}$ & $0.22501\pm 0.00068$ &\cite{ParticleDataGroup:2024cfk}
				&
				$\sin\theta_{13}$ & $0.003732^{+0.000090}_{-0.000085}$ &\cite{ParticleDataGroup:2024cfk}
				\\
				$\sin\theta_{23}$ & $0.04183^{+0.00079}_{-0.00069}$ &\cite{ParticleDataGroup:2024cfk}
				&
				$\delta$ & $1.147\pm 0.026$ &\cite{ParticleDataGroup:2024cfk}
				\\
				$\bar{m}_c(m_c)$ & $1.2730\pm0.0046$ &\cite{ParticleDataGroup:2024cfk}
				&
				$\bar{m}_b(m_b)$ & $4.183\pm 0.007 $ &\cite{ParticleDataGroup:2024cfk}
				\\
				$m_b^{\rm kin}$ & $4.573\pm 0.012$ &\cite{Bordone:2021oof}
				&
				$m_t^{\rm pole}$ & $172.4\pm0.7$ & \cite{ParticleDataGroup:2024cfk}
				\\
				$\mu_\pi^2$& $0.477\pm 0.056$ &\cite{Bordone:2021oof}
				&
				$\mu_G^2$ & $0.306\pm 0.050$ &\cite{Bordone:2021oof}
				\\
				$\bar{B}_1(\bar{m}_b)$ & $1.028^{+0.064}_{-0.056}$& \cite{Kirk:2017juj}
				&
				$\bar{B}_2(\bar{m}_b)$ & $0.988^{+0.087}_{-0.079}$ &\cite{Kirk:2017juj}
				\\
				$\bar{\epsilon}_1(\bar{m}_b)$ & $-0.107^{+0.028}_{-0.029}$ &\cite{Kirk:2017juj}
				&
				$\bar{\epsilon}_2(\bar{m}_b)$ & $-0.033\pm 0.021$& \cite{Kirk:2017juj}
				\\
				$B_1^d(\bar{m}_b)$& $0.835\pm 0.028$ & \cite{DiLuzio:2019jyq}
				&
				$B_2^d(\bar{m}_b)$& $0.791\pm 0.034$ &\cite{DiLuzio:2019jyq}
				\\
				$B_3^d(\bar{m}_b)$& $0.775\pm 0.054$& \cite{DiLuzio:2019jyq}
				&
				$B_4^d(\bar{m}_b)$& $1.063\pm 0.041$ &\cite{DiLuzio:2019jyq}
				\\
				$B_5^d(\bar{m}_b)$& $0.994\pm 0.037$ &\cite{DiLuzio:2019jyq}
				&
				$B_{R_2}^s$&  $0.89\pm 0.38$ &\cite{Davies:2019gnp}
				\\
				$B_{R_3}^s$&  $1.07\pm 0.42$ &\cite{Davies:2019gnp}
				&
				$G_F$& $1.1663788\times 10^{-5}$ &\cite{ParticleDataGroup:2024cfk}
				\\
				$f_{\pi^\pm}$ & $0.1302\pm 0.0008$ &\cite{FlavourLatticeAveragingGroupFLAG:2021npn}
				&
				$f_{K^\pm}$ & $0.1557\pm 0.0003$ &\cite{FlavourLatticeAveragingGroupFLAG:2021npn}
				\\
				$f_{D}$ & $0.2120\pm 0.0007$ &\cite{FlavourLatticeAveragingGroupFLAG:2021npn}
				&
				$f_{D_s}$ & $0.2499\pm 0.0005$ &\cite{FlavourLatticeAveragingGroupFLAG:2021npn}
				\\
				$f_{B}$ & $0.1900\pm0.0013$ &\cite{FlavourLatticeAveragingGroupFLAG:2021npn}
				&
				$f_{B_s}$ & $0.2303\pm 0.0013$ &\cite{FlavourLatticeAveragingGroupFLAG:2021npn}
				\\
				$f_{\rho}$ & $0.213\pm 0.005$ &\cite{Bharucha:2015bzk}
				&
				$f_{K^*}$ & $0.204\pm 0.007$ &\cite{Bharucha:2015bzk}
				\\
				$f_{D^*}$ & $0.242^{+0.020}_{-0.012}$ &\cite{Gelhausen:2013wia}
				&
				$f_{D^*_s}$ & $0.293^{+0.019}_{-0.014}$ &\cite{Gelhausen:2013wia}
				\\
				$F_0^{BD}(m_\pi^2)$ & $0.669\pm 0.010$ &\cite{Endo:2021ifc} 
				&
				$F_0^{BD}(m_K^2)$ &   $0.672\pm 0.010$ &\cite{Endo:2021ifc}
				\\
				$A_0^{BD^*}(m_\pi^2)$  & $0.725\pm 0.014$ &\cite{Endo:2021ifc} 
				&
				$A_0^{BD^*}(m_K^2)$ &  $0.732\pm 0.014$ &\cite{Endo:2021ifc}
				\\
				$F_+^{BD}(m_\rho^2)$ & $0.686\pm 0.010$ &\cite{Endo:2021ifc} 
				&
				$F_+^{BD}(m_{K^*}^2)$ & $0.692\pm 0.010$ &\cite{Endo:2021ifc}
				\\
				$F_0^{B_sK}(m_D^2)$ & $0.310$ & \cite{Endo:2021ifc}
				&
				$F_0^{B\pi}(m_D^2)$ & $0.288$ & \cite{Endo:2021ifc}
				\\
				$F_+^{B_sK}(m_{D^*}^2)$ & $0.357$ & \cite{Endo:2021ifc}
				&
				$F_+^{B\pi}(m_{D^*}^2)$ & $0.328$ & \cite{Endo:2021ifc}
				\\
				$A_0^{B_sK^*}(m_{D}^2)$ & $0.438$ & \cite{Endo:2021ifc}
				&
				$A_0^{B\rho}(m_{D}^2)$ & $0.432$ & \cite{Endo:2021ifc}
				\\
				$|V_{ud}|$ & $0.97367\pm 0.00032$ & \cite{ParticleDataGroup:2024cfk}
				&
				$|V_{us}|$ & $0.22431\pm 0.00085$ & \cite{ParticleDataGroup:2024cfk}
				\\
				$|V_{cb}|$ & $0.0397\pm 0.0006$ & \cite{Iguro:2020cpg}
				&&&
				\\
				\hline
			\end{tabular}
		\end{table}
	\end{center}

\end{appendices}
\let\clearpage\origclearpage

%\bibliography{apssamp}% Produces the bibliography via BibTeX.

\end{document}